\documentclass[acmsmall, nonacm]{acmart}
\setcopyright{none}

\usepackage{xcolor}
\usepackage{tcolorbox}

\newtcolorbox{myquote}{colback=blue!5!white, colframe=blue!75!black, rounded corners, boxrule=0.4mm}

\AtBeginDocument{%
  \providecommand\BibTeX{{%
    \normalfont B\kern-0.5em{\scshape i\kern-0.25em b}\kern-0.8em\TeX}}}

\begin{document}

\title{ChatCollab: Exploring Collaboration Between Humans and AI Agents in Software Teams}

\author{Benjamin Klieger}
\authornote{All authors have contributed equally to this research.}
\email{klieger@stanford.edu}
\orcid{0009-0001-5011-7982}
\affiliation{
  \institution{Stanford University}
  \city{Stanford}
  \state{CA}
  \country{USA}
  \postcode{94305}
}

\author{Charis Charitsis}
\email{charis@stanford.edu}
\orcid{0000-0001-6839-2205}
\authornotemark[1]
\affiliation{
  \institution{Stanford University}
  \city{Stanford}
  \state{CA}
  \country{USA}
  \postcode{94305}
}

\author{Miroslav Suzara}
\email{msuzara@stanford.edu}
\orcid{0000-0003-4552-7861}
\authornotemark[1]
\affiliation{
  \institution{Stanford University}
  \city{Stanford}
  \state{CA}
  \country{USA}
  \postcode{94305}
}

\author{Sierra Wang}
\email{sierraw@stanford.edu}
\orcid{0000-0003-1376-8759}
\authornotemark[1]
\affiliation{
  \institution{Stanford University}
  \city{Stanford}
  \state{CA}
  \country{USA}
  \postcode{94305}
}

\author{Nick Haber}
\email{nhaber@stanford.edu}
\orcid{0000-0001-8804-7804}
\authornotemark[1]
\affiliation{
  \institution{Stanford University}
  \city{Stanford}
  \state{CA}
  \country{USA}
  \postcode{94305}
}

\author{John C. Mitchell}
\email{jcm@stanford.edu}
\orcid{0000-0002-0024-860X}
\authornotemark[1]
\affiliation{
  \institution{Stanford University}
  \city{Stanford}
  \state{CA}
  \country{USA}
  \postcode{94305}
}

\renewcommand{\shortauthors}{Klieger, Charitsis, Suzara, Wang, Haber, Mitchell}
\newcommand{\jcm}[1]{\textcolor{blue}{JCM: #1}}

\begin{abstract}

\textit{Abstract}
We explore the potential for productive team-based collaboration between humans and Artificial Intelligence (AI) 
by presenting and conducting initial tests with a general framework that enables multiple human and AI agents to work together as peers. ChatCollab’s novel architecture allows agents - human or AI - to join collaborations in any role, autonomously engage in tasks and communication within Slack, and remain agnostic to whether their collaborators are human or AI. Using software engineering as a case study, we find that our AI agents successfully identify their roles and responsibilities, coordinate with other agents, and await requested inputs or deliverables before proceeding. In relation to three prior multi-agent AI systems for software development, we find ChatCollab AI agents produce comparable or better software in an interactive game development task. We also propose an automated method for analyzing collaboration dynamics that effectively identifies behavioral characteristics of agents with distinct roles, allowing us to quantitatively compare collaboration dynamics in a range of experimental conditions. For example, in comparing ChatCollab AI agents, we find that an AI CEO agent generally provides suggestions  2-4 times more often than an AI product manager or AI developer, suggesting agents within ChatCollab can meaningfully adopt differentiated collaborative roles. Our code and data can be found at: \url{https://github.com/ChatCollab}.
\end{abstract}



\maketitle

\section{Introduction}
\label{sec:intro}
Advancements in Large Language Models (LLMs) and Artificial Intelligence (AI) agent systems have significantly increased the potential for productive collaboration between humans and AI.
Tools like GitHub Copilot \cite{githubCopilot} have become widely used. However, popular copilots place humans in the primary decision-making role and typically do not alter the dynamics of human-to-human interaction.
Most prior research on human-AI collaboration to date (\textit{e.g.,} \cite{10.1145/3610054,10.1145/3579612,10.1145/3610218,10.1145/3610068,10.1145/3579601,10.1145/3610187,10.1145/3579476,10.1145/3610072,10.5555/3086747,doi:10.1177/0018720820960865})
has similarly focused on dyadic interaction and quantitatively measurable tasks such as collaborative gaming \cite{10.1145/3610072}, 
object identification \cite{10.1145/3491102.3517791} or decision making \cite{10.1145/3579476}. 
More recently, a new class of AI agent systems \cite{metaGPT1,ChatDev,Qian:arXiv2023,SuperAGI}
has emerged,  coordinating multiple AI agents to complete software development or similar tasks.
Looking ahead, however, we believe that more flexibly structured collaborative teams comprising both humans and AI agents
will become more prominent, allowing both humans and AI agents to assume diverse roles as needed. 
This is a natural evolution from structured human-only teams, allowing human-AI collaboration to draw on 
the rich history of research and practical experience in management and organizational behavior.

To explore the broad potential for productive team-based collaboration between humans and AI, we present ChatCollab, a configurable experimental system that supports AI agents as peer collaborators with humans. While ChatCollab is inherently task-agnostic, we have initially applied it to software development. Software teams offer a compelling use case due to the inherent complexity of the development process, the complementary nature of AI coding abilities to human skills, and the ability to leverage familiar team roles such as designers, engineers, product managers, and quality assurance testers.

In ChatCollab, one or more humans can actively participate as team members alongside any number of AI agents, each filling roles such as software developers, UI/UX designers, or quality assurance testers. 
Additionally, ChatCollab has the potential to facilitate human learning, particularly in contributing effectively to structured software teams. 
In human learning, AI agents with greater expertise can act as mentors, for example, while others can serve as practice partners \cite{socialskills,rehearsalAI} or teachable agents \cite{Biswas:AAI2005}.
This framework might help humans acquire new skills, refine their existing abilities, and better integrate into collaborative, AI-augmented environments.

We evaluate ChatCollab outcomes both by evaluating the quality of the software produced (work product) and by measuring characteristics of team interaction.
It is relatively straightforward to see that ChatCollab’s AI agents identify their roles and responsibilities, autonomously follow established processes, coordinate and communicate with other agents, and provide updates while appropriately waiting for input before proceeding. 
To evaluate collaborative interaction more deeply, we describe a method for AI-based collaboration analysis
that is based in part on traditional qualitative coding methods used in the social sciences (see section \ref{sec:collabAnalysis}).
Using this automated method,
we show how varying role definitions and social conventions (expressed as AI prompts) leads to measurable variations in the collaborative behavior of agents. 
We also measure ways that AI agents collaborate differently based upon their roles.
Looking at the quality of work product, we find that ChatCollab produces equal or superior software to three prior systems that use multiple AI agents for software development \cite{metaGPT1,ChatDev,SuperAGI}, while additionally allowing humans to engage directly in the development process. 

The main contributions of this paper include:
\begin{itemize}
    \item We introduce ChatCollab, a configurable system for human-AI collaborative teams, agnostic to application but used in this paper for software development. 
    \item We introduce and demonstrate a collaboration analysis method to measure the effectiveness of prompting AI agents to embody different roles and demonstrate specific social knowledge through collaborative behavior.
    \item We compare the characteristics of software produced by three prior multi-AI-agent systems, MetaGPT, ChatDev, and SuperAGI \cite{metaGPT1,ChatDev,SuperAGI}, all designed for software development, with each other and with ChatCollab in a case study. 
\end{itemize}

The paper is organized with section \ref{sec:related_work} summarizing the relevant background for the development of ChatCollab, section \ref{sec:chat_collab} introducing the ChatCollab system, section \ref{sec:collaboration_evaluation} detailing the collaboration analysis method, section \ref{sec:code_quality_evaluation} presenting the code quality evaluation, where we compare ChatCollab to three other multi-AI-agent systems, and  section \ref{sec:limitations} describing limitations of our system and our analyses.
Finally, section \ref{sec:conclusion} concludes and proposes directions for future work.

\section{Related Work}
\label{sec:related_work}

\subsection{AI-Assisted Software Development}

There is a rich history of AI-based tools for specific aspects of software development.
Summaries organized according to phases of the software development process appear in 
\cite{DBLP:journals/corr/abs-2108-01591,AppAIclassicalSE}.
Beginning with Github Copilot \cite{githubCopilot},
a number of copilot tools integrate calls to an LLM into the IDE.
A comparative study of Github Copilot, Tabnine, Replit Ghostwriter, and Codeium appears in a Codeium blog post \cite{AICodeAssistants2023}.
Research studies of the power and limitations of Github copilot and direct 
access to LLMs appear in \cite{DBLP:journals/corr/abs-2107-03374,10.1145/3633453}.

\subsection{Multi-AI-Agent Software Development}

Prior work observes that LLMs may struggle with hierarchical multi-step reasoning 
tasks like generating complex programs \cite{DBLP:journals/corr/abs-2107-03374}.
Observing this limitation, Zelikman et al. \cite{zelikman2023parsel} introduce a framework that automatically decomposes algorithmic tasks into hierarchical natural language function descriptions and then proceeds to select suitable code implementations for each function. 
The advantage of enlisting multiple agents is illustrated by the fact that in comparison to directly sampling AlphaCode and Codex,
the system developed by Zelikman et al.  solves more competition-level problems in the APPS dataset, resulting in over 75 percent higher pass rates. 

Prior systems have used multiple AI agents to simulate realistic human behavior \cite{park2023generative}
and, for collaborative software development in particular, have assigned specific roles to agents
\cite{Qian:arXiv2023, hong2023metagpt}. 
In particular, the three structured agent systems MetaGPT, ChatDev, and SuperAGI \cite{metaGPT1,ChatDev,SuperAGI} 
we compare to ChatCollab and to each other all divide tasks and responsibilities using familiar human roles. 

\subsection{Human interaction with AI agents for Learning}

We are not aware of extensive prior work on multi-agent systems in 
which multiple humans interact as peers with multiple agents to produce a work product.
However, some interesting prior systems explore 
educational settings that involve humans and multiple AI agents.
Many of them seem best regarded as learning through role-playing or simulation. 

A recent position paper explains and summarizes prior use of a partner-mentor paradigm,
in which one AI agent serves as a mentor (or instructor) and another AI agent serves as a practice partner \cite{socialskills}. 
Examples cited include
Active Listening, 
Conflict Avoidance,
Conflict Resolution,
Empathy, and
Rhetoric, as well
as interactive TA (teacher) training as described below; a compelling case study appears in \cite{rehearsalAI}.
Biswas et al. propose a learning system with a multi-agent architecture that included four agents: a teachable agent, a mentor agent, a student agent, and an environment agent \cite{Biswas:AAI2005}. Similarly, Soliman and G\"utl identify and distinguish between the types of Pedagogical Agents which contribute to learning in Virtual Learning Environments: agents for learning personalization, teachable agents, and multiple agents supporting group learning \cite{Soliman:ICICL2011}.
Markel et al. present GPTeach, in which a pair of student AI agents provide practice for interactive TA training \cite{Markel:L@S2023}. 

More broadly, educational simulations with technology have been shown to enhance learning in technical training \cite{kincaid2003simulation}, games \cite{eikaas2006game} and engineering education \cite{babich2009teaching}.  
Theoretical foundations for educational simulations have been explored in \cite{lyons2012learning}
and elsewhere.
In the area of AI that improve human-to-human communication, 
AI-based chat interventions may improve the cordiality of political conversation between two humans online \cite{Argyle:PNAS2023}. 
Similarly, human-AI collaboration through AI-in-the-loop has also been shown to increase conversational empathy \cite{Sharma:Nature2023}. 

\subsection{Collaboration analysis}
\label{sec:collabAnalysis}

We view our analysis of ChatCollab transcripts -- sequences of message between members of the team -- as an automated form of qualitative coding. 
Coding in social science research
is a qualitative data analysis method where descriptive labels are assigned to selected aspects of the data,
allowing researchers to tabulate qualitative features.
Coding labels are typically assigned by human coders who are given specific criteria on when to apply them.
There are a number of methods, including comparison of codes assigned by multiple coders to assess confidence and improve tabulation. 
Because human data labeling is often a bottleneck or resource-constrained limitation of social science research, there is a growing body of recent work aimed at automated coding using LLMs. Some of the main areas of investigation are summarized below.

\textbf{Qualitative coding using LLMs}
In one study, GPT-4 achieves human-like performance in several coding tasks, with significant improvement when chain-of-thought reasoning is used \cite{dunivin2024scalable}. 
A comparison of GPT-3.5 Turbo and GPT-4o for inductive qualitative coding indicate that GPT-4o outperformed GPT-3.5, especially at low temperatures, by offering higher agreement rates and more consistent results compared to human coders \cite{arlinghaus2024inductive}.
A related proposal may effectively assist or replace human coders by achieving comparable accuracy and reducing coding time \cite{chew2023llm}.
Coding interview transcriptions through generative coding and lexico-semantic coding to allow for semi-automated development of codebooks, enhances coding accuracy and thematic relevance while potentially reducing human bias and error
\cite{bryda2024words}.
A comparison using zero-shot, few-shot, and contextual coding on virtual tutoring session transcripts suggests that GPT-4 Turbo is generally effective for well-defined constructs but struggles with complex constructs and contextually dependent cases \cite{liuassessing}.
Another study presents an AI-based tool that assists with both inductive (data-driven) and deductive (codebook-based) coding, improving efficiency, consistency, and accessibility in the coding process; substantial alignment between human coders and QualiGPT is reported \cite{zhang2024qualitative}.

Looking at bias and alternative language contexts, a study of  ChatGPT and Llama-2 focusing on transcripts from 2,407 interviews with Rohingya refugees and Bangladeshi residents discussing their aspirations for their children, finds that LLMs introduce systematic biases, especially in context-sensitive data. These biases result in patterned, non-random errors; supervised models trained on smaller, high-quality human annotations yield more accurate and unbiased results \cite{ashwin2023using}.

\textbf{Semi-Automated qualitative coding}
Studying semi-automated qualitative coding, investigators found that researchers prefer automation only after creating a codebook and coding an initial data subset, especially for extending codes to unseen data. 
Their prototype tool using simple NLP techniques achieves human-comparable inter-rater reliability
\cite{marathe2018semi}.
Focusing on ambiguity, which often appears as coder disagreement, another proposed tool highlights areas of coder disagreement, allowing collaborative coders to explore inconsistencies and gain deeper insights
\cite{chen2018using}.
Additional tools  to systematically analyze, improve, and verify coder agreement
have also been proposed \cite{kuckartz2019analyzing}.
Drawing on their experience with 18 studies involving 12 to 54 coders, another effort
addresses pragmatic challenges such as recruiting and training coders, ensuring data quality and coding reliability at scale, and maintaining team cohesion \cite{beresford2022coding}.
\section{ChatCollab: Configurable System for Human-AI Collaboration}
\label{sec:chat_collab}

We present ChatCollab, a system that enables configurable teams composed of both humans and AI agents. 
Users define specific team roles assigned individually to each agent in the team --- human or AI --- allowing for flexible team configurations and enabling authentic collaboration between humans and AI agents as peers. 

\subsection{Motivation for ChatCollab design decisions}

\par 
AI-Human collaboration may be designed to produce a work product,
have some effect on the participants, or both. 
In a single sprint of a hybrid human-AI software team, for example, the primary goal may be the best software possible. However, if a team operates over a longer period of time, evolutionary improvements in the team are also important. 
Many team leaders and managers in human organizations recognize the 
importance of balancing work-produce goals with team morale, future 
productivity, and career development for members of the team.
Because environments that use multiple intelligent agents
to assist student learning address human engagement and progress, 
we have found it useful to draw on that literature, \textit{e.g.,} \cite{Biswas:AAI2005,Soliman:ICICL2011}.
This perspective led us to the following design goals:

\begin{itemize}
    \item AI agents embody different roles within the collaboration, acting different from one another rather than being monolithic, thus forming a genuine collaborative team.
    \item AI agents within the team can be prompted to behave meaningfully differently within the collaboration, offering humans the ability to steer the team dynamics produced by the system.
    \item The human(s) can assume any role within the AI Agent team, experiencing realistic collaboration. The human(s) can direct the team’s progress, or have it directed by an AI Agent.
    \item AI agents can be taught or give feedback by the human participants through direct, iterative interactions. The human can both coach and be coached by the AI Agents.
    \item The environment, including a shared workspace and the details of each agent’s role, is conveyed equally and fully to both the human and AI agents. 
\end{itemize}

\par In a software development team, the hierarchical structure naturally allows for the emergence of teachable agents, who accept instruction, and mentor agents, who provide guidance and encouragement. 
In practice, a human senior developer might provide feedback to an AI junior developer, while reporting to an AI CTO. 
This structure fosters the development of teamwork skills for human participants by engaging with agents in various capacities.

\par In order to achieve the goals outlined above,
we hypothesize that individual AI agents must be designed with the following features:

\begin{itemize}
    \item AI agents must have full autonomy, defined as the ability to take or not take action at any update in the environment, without following a predetermined sequence of decisions. This means the system must enable agents to be continuously aware of their environment and able to take or attempt actions at any time if the agent expresses such a desire.
    \item AI agents should possess sufficient intelligence to be able to learn from human instruction and provide helpful feedback.
    \item AI agents should have believable social abilities and environmental awareness, enabling realistic collaborations with human participants. This includes the ability to follow expected social etiquette in a software development team, both standardized and team-specific.
    \item AI agents should each have acccess to contextual information about their environment, their role in completing the task at hand, and their plans to complete those tasks.
    \item AI agents should have optional customizable characteristics beyond their roles in the team, including traits such as confidence, personality, knowledge level, and receptiveness. 
\end{itemize}

\par Autonomy is important to allow agents to fully respond to the human and other participants. Agents should be responsive to human direction, able to deviate from a sequence of artifact generation to communicate and collaborate with the human participants in a free-flowing dialogue. As mentor agents that might direct or manage humans, they must be able to provide spontaneous feedback to the human participant. 
The remaining requirements focus on the cognitive and social abilities of the agents. For agents to be teachable, they must be able to learn from human instruction. For agents to be supervisory, they must be capable of providing helpful feedback to humans. 
The cognitive requirements for the agents likely require their access to the environment, their role, and actions.

\subsection{ChatCollab System Description}

At a high level, the system works as follows:
\begin{enumerate}
    \item A human administrator defines roles of the team by entering them into an admin dashboard. 
    \item Each AI agent is created through the system and begins running autonomously.
    \item A shared event timeline is created that all agents have access to for communication and taking actions.
    \item At intervals, each AI agent checks the shared event timeline. 
    \begin{enumerate}
        \item If there is a new event, the agent can decide whether to take an action: compose a message, generate a file, or do nothing. 
        \item If the agent begins writing a message, an event will be added to the event timeline. This event will be added to the prompts of other agents so they know that another agent is typing. The individual agents can still decide to send a message, even though another agent is typing. 
        \item This cycle of event to decision to action or inaction powers the agents’ autonomy, as they are able to observe, make decisions, and take action at any time. 
        This is the same level of autonomy a human would have in a team.
    \end{enumerate}
\end{enumerate}

\begin{figure}
  \centering
  \includegraphics[width=\textwidth, height=3.5in]{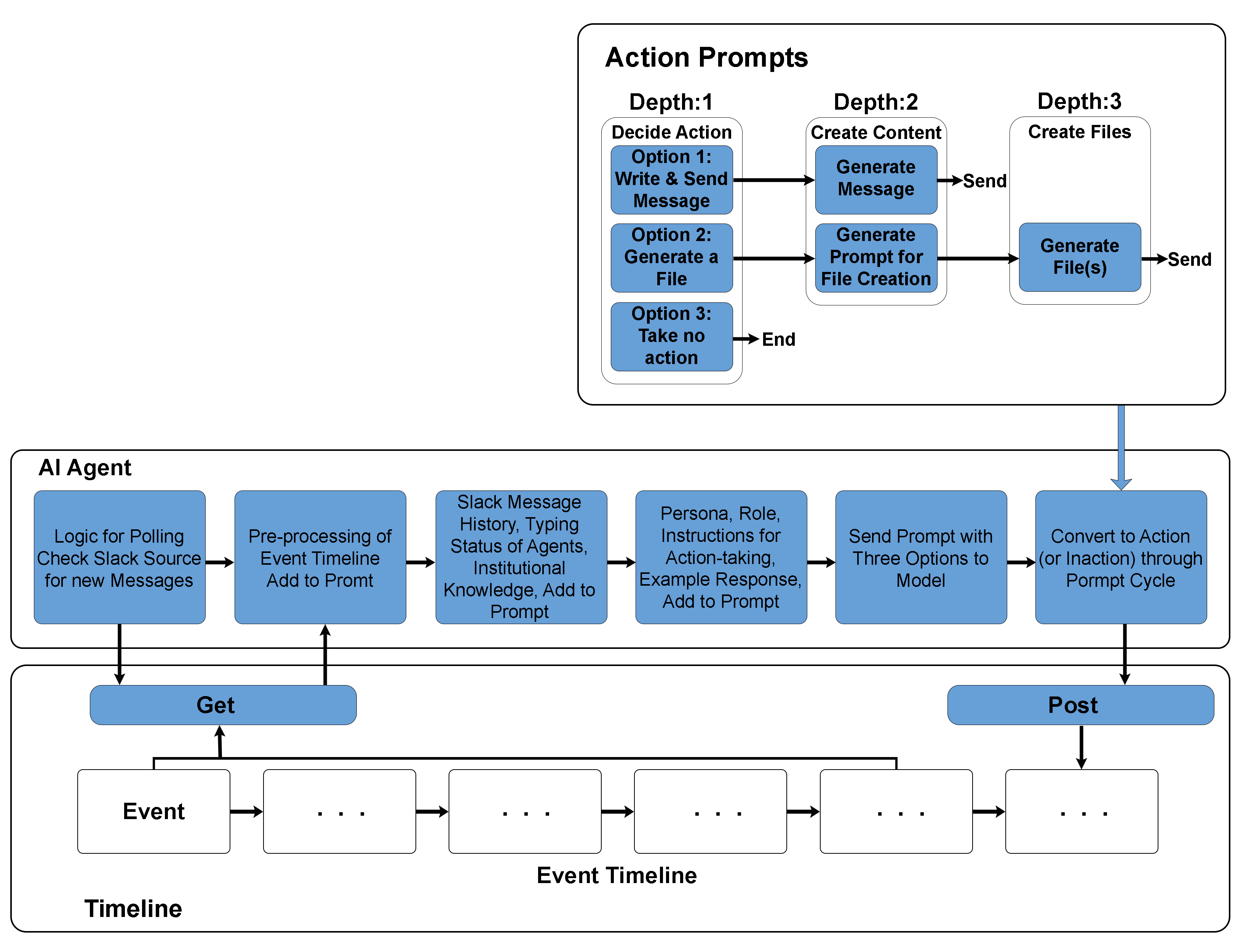}
  \caption{ChatCollab agent schematic and event timeline.
  }
  \label{fig:agent_framework__general_architecture}
\end{figure}

\begin{figure}[h]
  \centering
  \includegraphics[width=\textwidth]{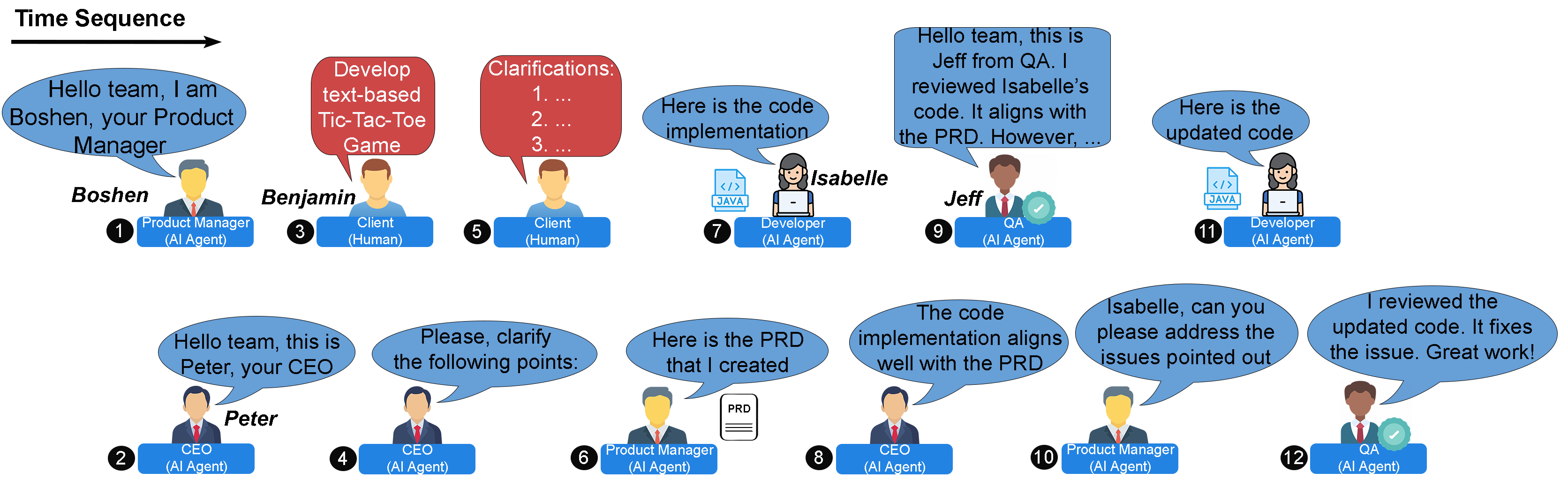}
  \caption{
  An exchange among a software development team of five members - a human client, an AI product manager, an AI CEO, an AI developer, and an AI QA - using ChatCollab.
  The official transcript is in \ref{appendix_human_ai_communication_transcripts}  
  }
  \Description{Description}
  \label{fig:message_exchange}
\end{figure}

\noindent Figure \ref{fig:agent_framework__general_architecture} illustrates this system. Figure \ref{fig:message_exchange} shows an example interaction of a run using ChatCollab.
 
\noindent We designed the ChatCollab system to meet our primary design goals in a number of significant ways:

\par \textbf{Shared event timeline and event driven actions.}
ChatCollab system uses an event-driven architecture for communication and actions. Events are organized into a centralized timeline. This creates a shared linear order for activity in the entire system, which ensures all agents can have the same level of information and visibility. The standardization of all action-taking and communication through the shared event timeline creates a common protocol for all agents. As all communication between AI agents occurs through the timeline, determining the requirements for replacing an AI agent with a human agent is simple. This approach also improves debugging capabilities, as it enables a human supervisor to view all communication and actions in a centralized location.

\textbf{Fully autonomous agents}.
Agents have the ability to choose to take any action at any time. However, their choices are governed by the prompted social character of the agents. For instance, in a scenario we explore, the product manager does not code because the underlying LLM knows that the role of a product manager is not to generate code. The developer similarly knows not to talk over the CEO because the underlying LLM understands the hierarchical structure of a software development team. A portion of the system description with captures collaborative conventions (or institutional knowledge) offers the opportunity to further clarify and influence collaborative behavior. 

\par \textbf{Customizable agents.}
The admin interface can be accessed by a human admin, allowing agents to be added at any point during the process. This can be used to reflect the property of human teams that new team members can be added during a project. AI agents also have customizable characteristics through the customizable role name and description they are given when initialized through the admin interface. 
The use of LLM prompts  offers considerable flexibility in simulating different scenarios within a team. For instance, AI agents can be given challenging personalities  or behaviors that require guidance, like a team member new to the field.

\par \textbf{All communication on Slack.}
All communication takes place in Slack. It is the only way agents communicate with each other. This means the human can see and participate in all communication that happens in the exact same manner that the AI agents communicate with each other: Slack messages in one shared group channel. A natural result is that swapping an AI agent with a human is as simple as having the human join the channel, rename themselves on Slack, and begin messaging. The human can talk to any of the AI agents at any time on a platform that is common for software development communication.

\par \textbf{User interface.} 
The internal reasoning produced by the ChatCollab AI agents when making decisions is made public to a human admin in the system's user interface.
The admin interface shows the internal reasoning for observability and allowing the human admin to engineer the institutional knowledge to develop configurable collaborative dynamics.

\textbf{Other design decisions}.
Several other design decisions were made to improve the capabilities of the AI agent team. Agents themselves are prompted to generate instructions for file generation, which creates a scalable approach as new document types and expectations are encountered in more complex or atypical interactions. In addition, the collaboration conventions, institutional knowledge, included in the system description shapes the team’s workflow.

\subsection{ChatCollab experiment methodology}

In our experiments, we configure a team of three AI agents and one human. This includes Peter (AI CEO), Boshen (AI Product Manager), Isabelle (AI Developer), and Benjamin (Human client). We experimented with other configurations, such as adding a QA testing agent or multiple software developer agents. We also had humans take on the different roles such as CEO, product manager, and developer in the system development process to confirm the system was functionally sufficient to allow a human to successfully collaborate with the team while in those roles. We changed agent personas to embody specific character traits such as frequently agreeable or frequently disagreeable. Finally, we also applied ChatCollab to domains outside of software engineering, such as tutoring a student with mathematics problems, to confirm the domain-agnostic nature of the system. 

In order to set a realistic scope of study and run experiments on prompting while holding the task and team structure constant, we focus the collaboration experimentation in this paper entirely on the specific tic-tac-toe case study with the original configuration of three AI agents and one human agent. Since the ChatCollab system is agnostic to whether an agent is human or AI, any participant in the team could be switched from AI to human or human to AI.

Each run was completed once the code was generated and all agents individually decided they had no further action to take. In our experiment, to standardize the human's behavior across runs, the same human participated in each of the experiments, and followed a specific playbook for when to interact. In each run, after 10 seconds, the human client greets all agents which have spoken. The human then provides the requirements and can answer any follow up clarification questions from the agents about them. Finally, if progress stalls, the human client can request a progress update with a concise message such as "Hello, what is the progress so far?". The full list of prompts, personas, and institutional knowledge is included in Appendix \ref{appendix_chatcollab_config_prompts}.

\subsection{Observed ChatCollab system behavior}

We find that in ChatCollab, the agents demonstrate awareness of social expectations and workflows in a software development team. The following examples, taken from a the transcript of interaction during development of our sample tic-tac-toe game by mutiple agents, illustrate how the agents are able to: (1) identify their roles and responsibilities, (2) follow established processes, (3) effectively coordinate and communicate, (4) acknowledge and update one another, and (5) wait for necessary events before proceeding.

\textbf{1. Role Identification and Responsibilities:} The agents try to clearly identify their roles and responsibilities within the team. For instance, Boshen, the Product Manager, outlines his role in the introduction to the team:
\begin{myquote}\small
\textbf{Boshen (AI Agent - Product Manager)}\newline
    ``Hello team, I'm Boshen, your Product Manager. I'll be working closely with you all to design efficient and user-friendly products. I'll be creating comprehensive PRDs post CEO's approval and will be available to answer any product-related queries. Looking forward to our collaboration.''
\end{myquote}

\textbf{2. Following Established Processes:} The agents exhibit a clear understanding of established processes in the team. For example, Isabelle, the Developer, recognizes the importance of waiting for the PRD before commencing development:
\begin{myquote}\small
\textbf{Isabelle (AI Agent - Developer)}\newline
    \textbf{[Internal Reasoning Log]} ``As Isabelle, the developer, I'm waiting for the PRD from the product manager, Boshen, before I can start working on the code. Since there's no PRD yet and no new messages or questions directed at me, there's no action needed from my side at this point.''
\end{myquote}

In this example, Isabelle decides to wait for the process to advance to a specific stage before taking action. This was not a preset behavior. Rather, the AI Agent Isabelle was prompted to make a decision about what action or inaction to take.
\newline

\textbf{3. Coordination and Communication:} Effective coordination and communication are emphasized throughout the interactions. The CEO, Peter, sets the tone for the project and ensures that all team members understand the steps and expectations:
\begin{myquote}\small
\textbf{Peter (AI Agent - CEO)}\newline
    ``Hello team, this is Peter, your CEO. I'm excited to start this new project with you all. Let's ensure we follow our process: clarifying questions to the client, generating the PRD, and then moving to development. Remember, strong documentation is key. Let's get started!''
\end{myquote}

\textbf{4. Acknowledging Messages and Providing Updates:} The agents frequently acknowledge messages and provide necessary updates as part of their workflow. This is illustrated when the Boshen, the Product Manager, welcomes Benjamin, the human client, and asks him for instructions:
\begin{myquote}\small
\textbf{Boshen (AI Agent - Product Manager)}\newline
    ``Hello Benjamin, welcome to the team! If you have any specific requirements or clarifications for the project, feel free to share. We're here to ensure the product meets your expectations.''
\end{myquote}

\textbf{5. Waiting for Input Before Proceeding:} The agents demonstrate the ability to wait for necessary input before proceeding with their tasks, which is crucial for avoiding redundant or misaligned work:
\begin{myquote}\small
\textbf{Boshen (AI Agent - Product Manager)}\newline
   \textbf{[Internal Reasoning Log]} ``The CEO has asked Benjamin (Client) some clarifying questions. As a product manager, I should wait for Benjamin's responses before proceeding with the PRD creation. This will ensure that the PRD accurately reflects the client's needs and expectations.''
\end{myquote}

These examples indicate that the agents not only recognize but can also effectively adhere to social expectations and standard practices within a software development team. This includes clear role definition, following processes, effective communication, acknowledging messages, and waiting for necessary inputs before taking further actions. These qualities were all recognized to be essential requirements for effective collaboration in a hybrid Human-AI agent team.
\section{Collaboration Analysis}
\label{sec:collaboration_evaluation}

To explore how AI agents within ChatCollab exhibit differing behaviors depending upon their role, we propose a method for classifying and analyzing the behavior of the AI Agents. In addition, this allows us to determine if AI agent collaborative behavior can be predictably changed through prompting of the institutional knowledge.
Our method leverages LLMs to
automatically classify sequences of agent actions according to any chosen classification framework.
We illustrate the method with a sample study using Bales’ Interaction Process Analysis Framework
\cite{Bales50}, a well-established method for categorizing group interactions.
While using LLMs for qualitative coding and categorization is not novel in itself (see section \ref{sec:related_work}), 
the current context provides a unique setting for evaluating collaborative dynamics. 

\subsection{Methodology}

To be clear, a ChatCollab \textit{run} is a sequence of actions carried out by agents, and a \textit{transcript} is the log of messages from a run.

\subsubsection*{Data Collection}
In the study reported here, we consider eight experimental conditions.
For each experimental condition, we conduct three runs of the ChatCollab system until completion of the code and record the full dialog transcript of AI and human agents for each run. This resulted in 237 transcript messages total across all runs, including the control.

\subsubsection*{Data Preparation and Cleaning}

For each run, ChatCollab outputs the conversational transcript with a pattern for displaying speakers, roles, timestamps, and messages to a markdown file.
We use a Python script to automate the extraction of conversational data from markdown files and convert it into structured CSV formats. 
Using regular expressions, the content of each file is split into dictionaries stored in a list, then written to a CSV file where each row represented a single conversational turn. 
This format facilitates easier analysis of turn-based interactions. 
The code for this script appears in Appendix A.

\subsubsection*{Framework Selection}
We chose to use the well established Bales’ Interaction Process Analysis Framework \cite{Bales50} to perform our collaboration analysis, though we believe our method is generalizable to other frameworks. The framework is as follows:
\begin{enumerate}
    \item Shows Solidarity: raises other’s status, gives help, reward.
    \item Shows Tension Release: jokes, laughs, shows satisfaction.
    \item Agrees: shows passive acceptance, understands, concurs, complies.
    \item Gives Suggestion: direction, implying autonomy for other.
    \item Gives Opinion: evaluation, analysis, expresses feeling, wish.
    \item Gives Orientation: information, repeats, clarifies, confirms.
    \item Asks for Orientation: information, repetition, confirmation.
    \item Asks for Opinion: evaluation, analysis, expression of feeling.
    \item Asks for Suggestion: direction, possible ways of action.
    \item Disagrees: shows passive rejection, formality, withholds help.
    \item Shows Tension: asks for help, withdraws out of field.
    \item Shows Antagonism: deflates other’s status, defends or asserts self.
\end{enumerate}

\subsubsection*{API Calls and Prompt Engineering}
We found it effective to use a straightforward prompt that included a full description of the task and the expected format of the response. We integrated Bales’ IPA framework into the prompt used for automated classification, using an iterative prompt development process with human verification to identify the optimal wording.
We then systematically applied this framework to classify each agent action using an LLM (OpenAI's GPT-4 Turbo). 
Note that results were single-coded to the best fitting category. 
We experimented with different lengths of surrounding context, ranging from the prior two turns to the full context from the given run. We found it best to keep the prompts concise.

The final prompt includes the main categorization criteria plus \textit{None of the Above} to account for cases that do not clearly fit into any of the categories. 
We provided the LLM with the roles and messages of up to two turns before, the current message itself, and two turns after the intended message. 
The prompt is provided in Appendix \ref{appendix_llm_labeling_prompt}.

\subsubsection*{Human Coding}
A portion of the data was additionally coded using human review from members of our team to check for accuracy. This was done to ensure the reliability of the LLM-augmented classification processes and allowed for iterative refinement of the prompts. 

\subsubsection*{Quantitative Analysis}
Once automatically coded by the LLM, we summed the counts of each interaction type across the conditions. 
This identified patterns of collaboration that might be influenced by the experimental conditions. 
We calculated frequencies and proportions of each interaction type for each experimental condition. 
This allowed us to examine the impact that the experimental conditions had on counts of collaborative interaction types.

\subsection{Results}

In our analysis, we investigated the interaction patterns and role dynamics across different experimental conditions. 
We wanted to understand 
(1) how the AI agents' role prompts in ChatCollab would impact their behavior, (2) if prompting within the institutional knowledge could predictably influence the collaborative dynamics of the agents, and 
(3) how effectively our analysis method would detect such effects.

\subsubsection*{Human-AI Coding Alignment}
We conducted a structured comparison between the LLM categorizations against human-coded categorizations applied with the same collaborative framework.

To assess the accuracy of automated coding for our specific experiment, we calculated inter-rater reliability metrics between the LLM and human codes. 
This resulted in pairwise percent agreement as 78.1\% and Cohen's Kappa as 72.9\%, which indicated significant alignment between human and AI coding.

\subsubsection*{Role-Specific Contributions}

\begin{figure}[h]
  \centering
  \includegraphics[width=\textwidth]{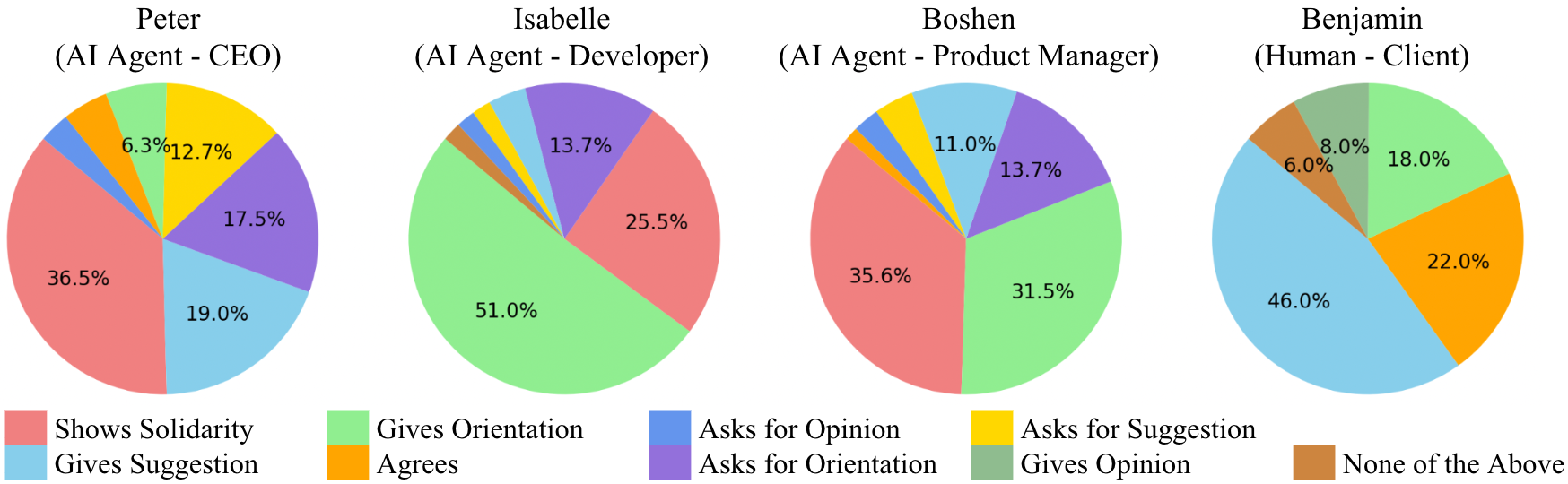}
  \caption{Each pie chart shows the distribution of message types sent by each agent in a ChatCollab run.}
  \Description{Role-based sequences across experimental conditions, showing the flow of messages and contributions by CEO, Product Manager, Developer, and Client.}
  \label{fig:role_pie_charts}
\end{figure}

Figure \ref{fig:role_pie_charts} shows the distribution of interaction moves made by each AI agent and the human client across all experimental conditions and runs. The manner in which the collaborative move for each run was prompted in institutional knowledge is included in Appendix \ref{appendix_chatcollab_config_prompts}.
This visualization provides insight on the distinct roles performed by each agent, reflecting alignment with expectations of their defined roles from the ChatCollab framework. 
Peter, the AI CEO, demonstrated an emphasis on "Shows Solidarity" and "Gives Suggestions". 
Isabelle, the AI developer, mostly engaged in "Gives Orientation" and "Shows Solidarity". Boshen, the AI Product Manager displayed a dual focus on both "Shows Solidarity" and "Gives Orientation". Benjamin, the human client, predominantly engaged in "Gives Suggestion".

Thus, this provides evidence suggesting agents within the system act meaningfully different depending upon their roles. For instance, Peter in the role of CEO gives suggestions 19.0\% of the time, compared to 11.0\% for Boshen the product manager and <5.0\% for Isabelle the developer. Similarly differentiated trends are seen for "Gives Orientation" and "Asks for Suggestion".

\subsubsection*{Collaborative Configurability}

We also examine the extent to which collaboration within teams of human-AI agents may be meaningfully guided through prompts that suggest specific interaction behavior. 
By comparing each experimental condition to a control group, we analyzed the percentage differences in interaction categories to understand how prompts influenced collaboration patterns. 
In this analysis we combined "Asks for" and "Gives" categories for Suggestion, Orientation, and Opinions. 
This decision was based on observing that these behaviors often function as reciprocal parts of the same interaction process - for example, when AI agents were prompted to ask for suggestions, we would expect that other agents reciprocate by giving suggestions as well. 
Collapsing these categories allowed us to capture a more holistic view of each prompt's influence on collaborative behavior just like the back-and-forth dialogue of team interactions.

\begin{figure}[h]
  \centering
  \includegraphics[width=\textwidth]{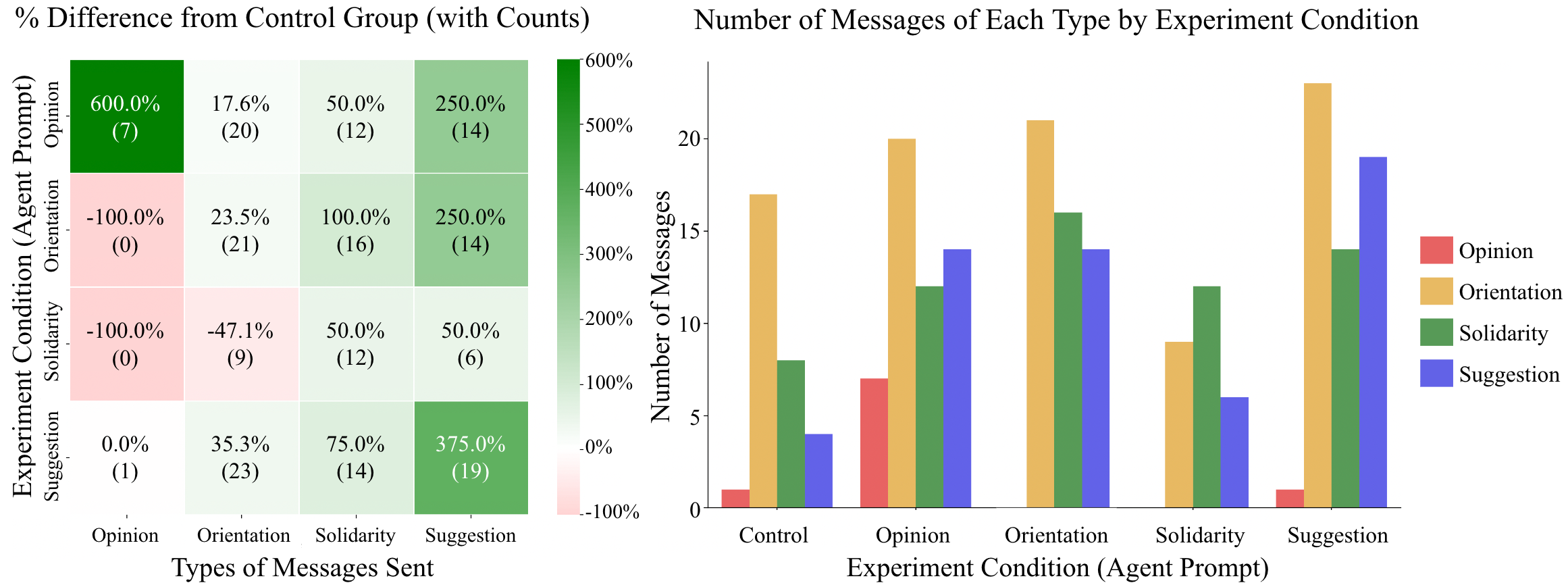}
  \caption{
  The heatmap (left) shows the percentage differences in message types across experimental conditions relative to the control condition, with raw message counts in parentheses. The bar chart (right) displays the raw counts of each message type for each experimental condition.
  }
\Description{Percentage difference in interaction categories across experimental conditions.}
\label{fig:conditions_figure_new}
\end{figure}

Tabulating the results shown in Figure~\ref{fig:conditions_figure_new}, we observe the following distinct outcomes:

\begin{enumerate}
    \item Gives/Asks for Opinion: Increase of +600\% compared to the control. This suggests that the prompt effectively guided the team toward giving and asking for opinions.
    \item Gives/Asks for Suggestion: Increase of +375\% compared to the control. This indicated that the prompt successfully encouraged participants to both ask for and offer advice.
    \item Gives/Asks for Orientation: Increase of 23.5\% compared to the control. This rise also increased Gives/Asks for Suggestion by +250\% and Shows Solidarity by 100\% which may indicate more context provision to collaborators.
    \item Shows Solidarity: Increase of +50\% in Shows Solidarity compared to the control. Interestingly there was a decrease of -47.1\% in Gives/Asks for Orientation compared to the control. However, because of the nature of single coded replies, it may be such that messages had dual-intent in giving orientation but grounded within shows of solidarity.
\end{enumerate}

These results demonstrate that ChatCollab prompts which target specific interaction categories can configure team behavior within collaborative AI-human environments. 
This suggests that collaborative dynamics in multi-AI-mediated interactions can be influenced by intentional prompting, where setting a clear behavioral focus leads to a measurable increase in that behavior within team interactions. This highlights the potential for ChatCollab to be used as a means to guide multi-AI collaboration in such a way that lines up with desired communication/collaboration patterns, processes, and goals.

\subsubsection*{Interaction Categories Across Conditions}
In addition to analyzing the collaboration changes that occurred across conditions, we also explored the interaction categories sequentially to understand how and when different collaborative interactions took place within a run. 
We found overall trends such as "Gives Suggestion" tending to appear in the early and middle segments of a run and "Gives Orientation" tending to appear from middle to the end segments of a run.
This is shown in the Appendix \ref{fig:interaction_sequences}. 

\section{Code Analysis}
\label{sec:code_quality_evaluation}

\par In order for these agent systems to be effective tools for software development and long term human-AI collaboration, their code output must be functional and high quality. 
We compared the efficacy of ChatCollab to ChatGPT and several other AI agent systems (SuperAGI, MetaGPT, ChatDev) to confirm that it is an adequate supporting environment for multi-agent software development, at least as
good as other systems previously designed specifically for this purpose. We used GPT-4 with at least ten runs for each system. The human was only the client in all runs, with an AI CEO, product manager, and developer. Some runs for the code quality experiment included an AI quality assurance agent.

\subsection{Benchmark Task Selection}
We opted for the relatively straightforward design of a text-based tic-tac-toe game for two main reasons. First, it is easier to clarify the expectations and requirements and, therefore, introduce metrics that favor objective comparison. Second, if we can find an example that is fairly simple to reason about and interpret but that nevertheless leads to failures, we might hope to extrapolate failure modes to more complex tasks.
We also considered other options, including Sudoku, but ultimately determined that tic-tac-toe presented a more clearly defined ruleset and lent itself to more straightforward objective evaluation.

\subsection{Prompt Formulation}
Formulating the proper prompt is crucial in our comparison. Intuitively, a short prompt like “\emph{write a program to play a tic-tac-toe game}” may seem sufficient. There are many parameters that need clarification, especially for establishing a standard for comparison across AI agent systems. 

In our prompt, we list a set of functional requirements that a software developer needs to be aware of and the target programming language (Java), but did not specify the implementation details as this is normally a developer's responsibility. 
Our final prompt can be found in the Appendix \ref{appendix_user_prompt}.

\subsection{Evaluation Criteria}
\label{evaluation_criteria}

\begin{table}[b]
  \caption{The criteria we used for the system comparison are geared towards consistency and objectivity to deliver a fair evaluation as much as possible.}
  \centering
  \begin{tabular}{l}
  \toprule  
  \multicolumn 1{c}{Functionality} \\
  1. The code compiles without errors\\
  2. Uses `X' and `O' for the two players\\
  3. Creates a 3x3 grid\\
  4. Guides the players through the game\\
  5. Starts the game by displaying an empty board\\
  6. Starts the game by assigning `X' to the first player and `O' to the second player\\
  7. Prompts the players to input their moves by specifying the row and column\\
  8. Handles non-integer input\\
  9. Ensures that the user input is not out-of-range\\
  10. Ensures that the user input is not in an already occupied cell\\
  11. Correct placement of X's and O's according to user input coordinates\\
  12. Displays the updated board after each move\\
  13 .Displays the final board after the game ends\\
  14 .Detects the winner\\
  15. Announces the result of game as soon as a player wins\\
  16. Announces the result of game if it ends in a tie\\
  17. After the game concludes, asks for new game and if so restarts the game\\
  \midrule
  \multicolumn 1{c}{Code Quality} \\
  1. Decomposition\\
  2. Source Code Documentation (general comments, inline comments, etc.)\\
  3. Supporting material (user instructions, summary, notes, etc.)\\
  \bottomrule  
  \end{tabular}
  \label{table:evaluation_criteria}  
\end{table}

\par Although there is inherent subjectivity in assessing code quality and other qualitative software development factors, we aimed to establish fair standards by quantifying aspects like human-agent interaction and functional performance. 
We established a set of measurable standards to reduce variability in our assessments, found in Table \ref{table:evaluation_criteria}. 

\subsection{Results}
\label{sec:results}

\begin{figure}[h]
  \centering
  \includegraphics[width=\textwidth]{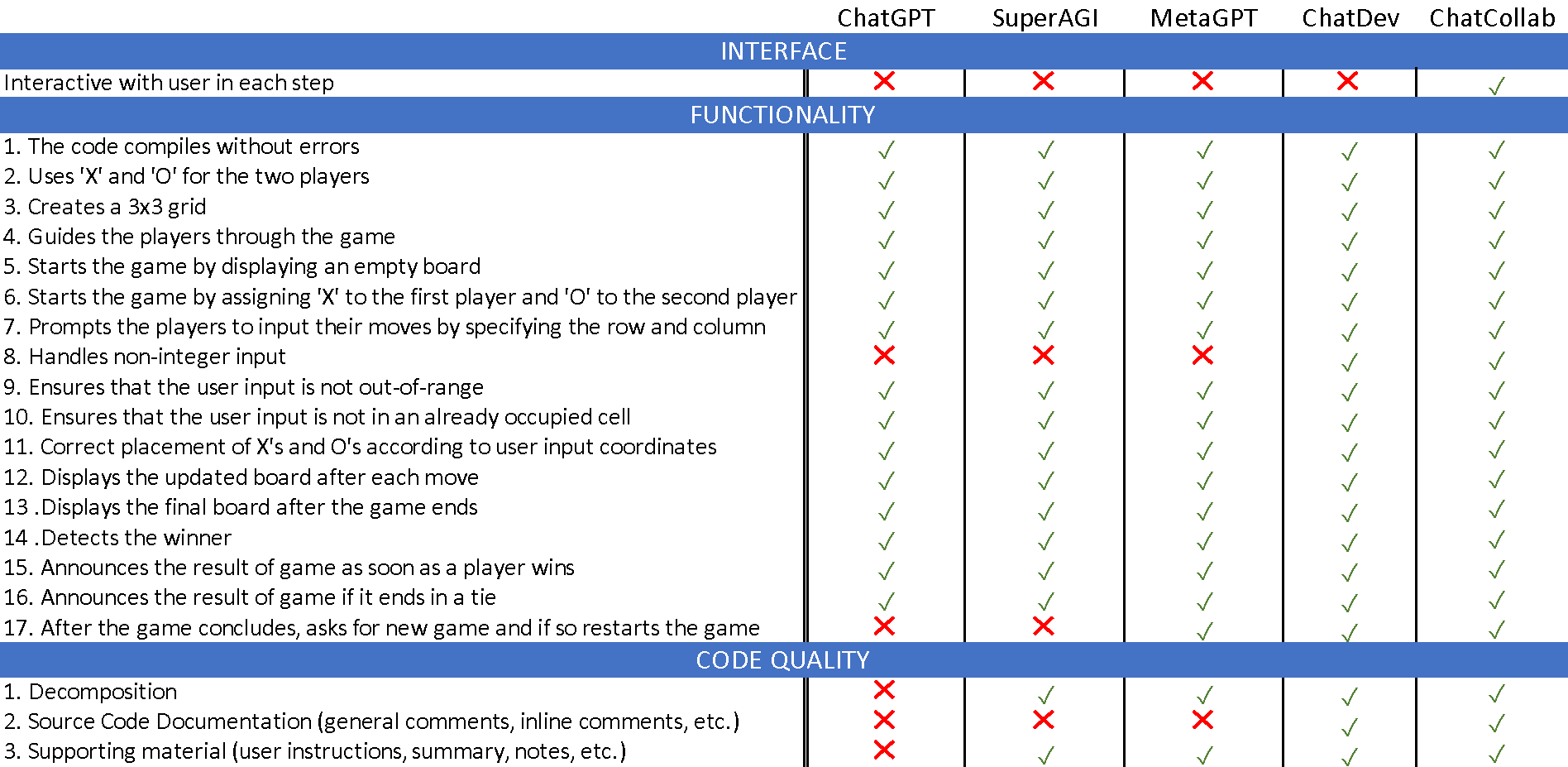}
  \caption{Comparison between LLMs, AI Agents and ChatCollab based on the criteria listed in Table \ref{table:evaluation_criteria}.}
  \Description{Comparison table between popular LLMs, AI Agents and ChatCollab}
  \label{fig:comparison_table}
\end{figure}

We evaluated several runs tasking agents in ChatCollab with producing the tic-tac-toe game, and observed the output quality of the code to be equal to or better than existing frameworks. Specifically, ChatCollab particularly excelled at feature inclusion and documentation because individual agents were able to proofread the code and request revisions to include anything missing. In addition, the inclusion of a product manager that asks clarifying questions to the human client allowed for improved accuracy of functionality.
Figure \ref{fig:comparison_table} summarizes how ChatCollab compares with popular LLMs and AI agent systems.
An example code output from a ChatCollab run with five AI agents, including specialized QA agents, and a human as the client is provided in the Appendix \ref{appendix_output_code_5_agents_and_human}.
\section{Limitations}
\label{sec:limitations}

This work is an initial exploration of multi-agent collaboration with inherent complexities. 
Limitations in the work and areas for future improvement are summarized below.

\textbf{Limited Collaboration Analysis}: 
We evaluated circumstances with up to five AI agents with different roles, and one human in the roles of CEO, developer, and product manager in different runs. 

It is possible that the system’s performance may vary as the complexity of scenarios increases, suggesting a need for further investigation to understand its scalability and robustness in more complex settings.
Future work may also consider multi-coded categories for message labeling, allowing a given message to support multiple categories within a collaborative framework rather than confined to one. Similarly, turns might be further deconstructed into multiple parts to break down individual meaning and address cases where a message labeled as Shows Solidarity also contained elements of Gives Orientation.

\textbf{Limited Code Quality Analysis}: Our code quality evaluation is based on a limited number of examples.
We have not conducted large-scale evaluation of human-AI collaboration in which we solicit others to participate in synthetic teams of various organization and size, for example. 
We hope, however, that by documenting our work so far, we are setting the stage for larger tests in future work.

\textbf{Biases of AI}: If this system is to gain widespread adoption across both industry and academic settings, it is important to meticulously evaluate its potential negative impacts on humans involved in the collaboration. It is known that AI can amplify the biases embedded within its training data. 
Further evaluation and mitigation are merited over time.

\textbf{Unknown Social Impact}: We have not yet conducted a dedicated user study with further investigation into the system’s impact on social dynamics and user experience. To ensure that this system is able to foster positive collaboration, we need to explore its effects on team dynamics, user interactions, and overall productivity. 

\textbf{Significant Computational Expenses}: ChatCollab has the potential for significant computational expenses compared to other systems. This is because all agents are able to be prompted after a new event, rather than a linear sequence of agents prompted. Several optimizations were made to reduce unnecessary token usage. For instance, agents are only prompted after a new event has occurred. This prevents duplicate prompting. The individual agents also experience a brief period of a random pause (3-15 seconds) between decisions to take an action, which reduces frequency of prompting and allows more space for humans to contribute to the conversation.
An alternative approach to a randomized pause could also be used to produce greater standardization of results.
Moving forward, more optimizations should be explored to reduce token usage and frequency. 

\textbf{Focus on Slack}: Although Slack is sufficient to produce product documentation, code, and a process which a human can observe and participate in, the sole usage of Slack without additional apps or platforms limits the functionality of ChatCollab. 
Future work should expand the platforms that are used for collaboration.

\textbf{AI Agents Don't Know Who is a Human}: The system may be limited by the agents’ lack of knowledge regarding which team members are human and which are AI. When agents are unaware of whether team members are human or AI, AI agents consistently treat all team members the same, preventing any unexpected changes in behavior when interacting with human members. However, it may create a limitation on learning and collaboration. The AI agent system could prioritize the learning experience of the human in the team, if they know which team member is a human. It may also aid the outcome of the collaboration. It should be noted that it is currently possible to identify the humans in the team through institutional knowledge, though this was not explored in our tests.

\textbf{Dependence on LLMs}: We assume that LLM performance will improve over time, and thus, the capabilities of the agents will similarly expand. 
It is conceivable that if there becomes a plateau in the performance of the agents, attributable to the limits of LLM advancements, it may become beneficial to implement hard-coded prompts and structures to further enhance their performance beyond this threshold. 
However, this would reduce the flexibility of the agent system, and thus, has been avoided by our new system architecture.

\section{Conclusion and Future Work}
\label{sec:conclusion}
In this paper, we introduce ChatCollab, a novel framework for AI-human collaboration, designed to explore the potential of configurable hybrid teams. 
ChatCollab enables dynamic role adoption and interaction between human and AI agents, without distinction based on the nature of the agent. 
This flexibility not only fosters peer collaboration, but also allows for the simulation of wide variety
of team environments. 

To assess the effectiveness of ChatCollab's prompting, we developed an automated method for analyzing collaboration dynamics,
present an illustrative case study, and conducted a benchmark code quality evaluation. 
Our results indicate that ChatCollab effectively modulates collaboration dynamics, in ways that can be measured 
using our collaboration analysis method. 
Furthermore, our benchmarks confirm that the code quality produced by ChatCollab is comparable to that of leading multi-agent systems, 
suggesting its viability as an effective software development system.

Moving forward, we are excited to explore the application of ChatCollab to a variety of collaborative tasks.
Because it is easy to configure using natural language prompting, a wide variety of experiments 
such as those described in \cite{socialskills} are easily implemented and conducted. The system can also expand beyond creating small teams and into larger organizational settings with human resources and people (or AI agent) managers.
We look forward to investigating how this system can be adapted to broader contexts
and used to study productivity and the quality of interactions within hybrid teams, including human-user studies to investigate impact on humans.
We also invite others to leverage ChatCollab for any possible experiments or applications for which it may be suited. The source code and data are publicly available on GitHub \cite{ChatCollab}.

\bibliographystyle{ACM-Reference-Format}

\begin{thebibliography}{48}


\ifx \showCODEN    \undefined \def \showCODEN     #1{\unskip}     \fi
\ifx \showDOI      \undefined \def \showDOI       #1{#1}\fi
\ifx \showISBNx    \undefined \def \showISBNx     #1{\unskip}     \fi
\ifx \showISBNxiii \undefined \def \showISBNxiii  #1{\unskip}     \fi
\ifx \showISSN     \undefined \def \showISSN      #1{\unskip}     \fi
\ifx \showLCCN     \undefined \def \showLCCN      #1{\unskip}     \fi
\ifx \shownote     \undefined \def \shownote      #1{#1}          \fi
\ifx \showarticletitle \undefined \def \showarticletitle #1{#1}   \fi
\ifx \showURL      \undefined \def \showURL       {\relax}        \fi
\providecommand\bibfield[2]{#2}
\providecommand\bibinfo[2]{#2}
\providecommand\natexlab[1]{#1}
\providecommand\showeprint[2][]{arXiv:#2}

\bibitem[git(2022)]%
        {githubCopilot}
 \bibinfo{year}{2022}\natexlab{}.
\newblock \bibinfo{title}{Github Copilot}.
\newblock \bibinfo{howpublished}{\url{https://github.com/features/copilot}}.
\newblock


\bibitem[Cha(2023)]%
        {ChatDev}
 \bibinfo{year}{2023}\natexlab{}.
\newblock \bibinfo{title}{ChatDev: Communicative Agents for Software Development}.
\newblock \bibinfo{howpublished}{\url{https://github.com/OpenBMB/ChatDev}}.
\newblock


\bibitem[Sup(2023)]%
        {SuperAGI}
 \bibinfo{year}{2023}\natexlab{}.
\newblock \bibinfo{title}{Introduction to SuperAGI}.
\newblock \bibinfo{howpublished}{\url{https://superagi.com/docs/}}.
\newblock


\bibitem[met(2023)]%
        {metaGPT1}
 \bibinfo{year}{2023}\natexlab{}.
\newblock \bibinfo{title}{MetaGPT github repository}.
\newblock \bibinfo{howpublished}{\url{https://github.com/geekan/MetaGPT}}.
\newblock


\bibitem[Cha(2024)]%
        {ChatCollab}
 \bibinfo{year}{2024}\natexlab{}.
\newblock \bibinfo{title}{ChatCollab's GitHub repository (Note: The non-anonymized GitHub link will be provided upon publication)}.
\newblock \bibinfo{howpublished}{\url{https://github.com/(anonymized)/ChatCollab}}.
\newblock


\bibitem[Argyle et~al\mbox{.}(2023)]%
        {Argyle:PNAS2023}
\bibfield{author}{\bibinfo{person}{Lisa~P. Argyle}, \bibinfo{person}{Christopher~A. Bail}, \bibinfo{person}{Ethan~C. Busby}, \bibinfo{person}{Joshua~R. Gubler}, \bibinfo{person}{Thomas Howe}, \bibinfo{person}{Christopher Rytting}, \bibinfo{person}{Taylor Sorensen}, {and} \bibinfo{person}{David Wingate}.} \bibinfo{year}{2023}\natexlab{}.
\newblock \showarticletitle{Leveraging AI for democratic discourse: Chat interventions can improve online political conversations at scale}.
\newblock \bibinfo{journal}{\emph{Proceedings of the National Academy of Sciences}} \bibinfo{volume}{120}, \bibinfo{number}{41} (\bibinfo{year}{2023}), \bibinfo{pages}{e2311627120}.
\newblock
\urldef\tempurl%
\url{https://doi.org/10.1073/pnas.2311627120}
\showDOI{\tempurl}
\showeprint{https://www.pnas.org/doi/pdf/10.1073/pnas.2311627120}


\bibitem[Arlinghaus et~al\mbox{.}(2024)]%
        {arlinghaus2024inductive}
\bibfield{author}{\bibinfo{person}{Clarissa~Sabrina Arlinghaus}, \bibinfo{person}{Charlotte Wulff}, \bibinfo{person}{G{\""u}nter~W Maier}, \bibinfo{person}{CS Arlinghaus}, \bibinfo{person}{C Wulff}, {and} \bibinfo{person}{GW Maier}.} \bibinfo{year}{2024}\natexlab{}.
\newblock \bibinfo{title}{Inductive Coding with ChatGPT-An Evaluation of Different GPT Models Clustering Qualitative Data into Categories}.
\newblock
\newblock


\bibitem[Ashwin et~al\mbox{.}(2023)]%
        {ashwin2023using}
\bibfield{author}{\bibinfo{person}{Julian Ashwin}, \bibinfo{person}{Aditya Chhabra}, {and} \bibinfo{person}{Vijayendra Rao}.} \bibinfo{year}{2023}\natexlab{}.
\newblock \showarticletitle{Using large language models for qualitative analysis can introduce serious bias}.
\newblock \bibinfo{journal}{\emph{arXiv preprint arXiv:2309.17147}} (\bibinfo{year}{2023}).
\newblock


\bibitem[Babich and Mavrommatis(2009)]%
        {babich2009teaching}
\bibfield{author}{\bibinfo{person}{Alexander Babich} {and} \bibinfo{person}{K~Th Mavrommatis}.} \bibinfo{year}{2009}\natexlab{}.
\newblock \showarticletitle{Teaching of complex technological processes using simulations}.
\newblock \bibinfo{journal}{\emph{The International journal of engineering education}} \bibinfo{volume}{25}, \bibinfo{number}{2} (\bibinfo{year}{2009}), \bibinfo{pages}{209--220}.
\newblock


\bibitem[Bales(1950)]%
        {Bales50}
\bibfield{author}{\bibinfo{person}{R.~F. Bales}.} \bibinfo{year}{1950}\natexlab{}.
\newblock \bibinfo{booktitle}{\emph{Interaction process analysis; a method for the study of small groups}}.
\newblock \bibinfo{publisher}{Addison-Wesley}.
\newblock


\bibitem[Barenkamp et~al\mbox{.}(2020)]%
        {AppAIclassicalSE}
\bibfield{author}{\bibinfo{person}{M. Barenkamp}, \bibinfo{person}{J. Rebstadt}, {and} \bibinfo{person}{O. Thomas}.} \bibinfo{year}{2020}\natexlab{}.
\newblock \showarticletitle{Applications of AI in classical software engineering}.
\newblock \bibinfo{journal}{\emph{AI Perspect}}  \bibinfo{volume}{2, 1} (\bibinfo{year}{2020}).
\newblock
\urldef\tempurl%
\url{https://doi.org/10.1186/s42467-020-00005-4}
\showURL{%
\tempurl}


\bibitem[Batarseh et~al\mbox{.}(2021)]%
        {DBLP:journals/corr/abs-2108-01591}
\bibfield{author}{\bibinfo{person}{F.A. Batarseh}, \bibinfo{person}{R.Mohod}, \bibinfo{person}{A. Kumar}, {and} \bibinfo{person}{J. Bui}.} \bibinfo{year}{2021}\natexlab{}.
\newblock \showarticletitle{The application of artificial intelligence in software engineering: a review challenging conventional wisdom}.
\newblock \bibinfo{journal}{\emph{CoRR}}  \bibinfo{volume}{abs/2108.01591} (\bibinfo{year}{2021}).
\newblock
\urldef\tempurl%
\url{https://arxiv.org/abs/2108.01591}
\showURL{%
\tempurl}


\bibitem[Beresford et~al\mbox{.}(2022)]%
        {beresford2022coding}
\bibfield{author}{\bibinfo{person}{Melissa Beresford}, \bibinfo{person}{Amber Wutich}, \bibinfo{person}{Margaret~V du Bray}, \bibinfo{person}{Alissa Ruth}, \bibinfo{person}{Rhian Stotts}, \bibinfo{person}{Cindi SturtzSreetharan}, {and} \bibinfo{person}{Alexandra Brewis}.} \bibinfo{year}{2022}\natexlab{}.
\newblock \showarticletitle{Coding qualitative data at scale: Guidance for large coder teams based on 18 studies}.
\newblock \bibinfo{journal}{\emph{International Journal of Qualitative Methods}}  \bibinfo{volume}{21} (\bibinfo{year}{2022}), \bibinfo{pages}{16094069221075860}.
\newblock


\bibitem[Biswas et~al\mbox{.}(2005)]%
        {Biswas:AAI2005}
\bibfield{author}{\bibinfo{person}{Gautam Biswas}, \bibinfo{person}{Krittaya Leelawong}, \bibinfo{person}{Daniel Schwartz}, \bibinfo{person}{Nancy Vye}, {and} \bibinfo{person}{The Teachable Agents~Group at Vanderbilt}.} \bibinfo{year}{2005}\natexlab{}.
\newblock \showarticletitle{Learning By Teaching: A New Agent Paradigm For Educational Software}.
\newblock \bibinfo{journal}{\emph{Applied Artificial Intelligence}} \bibinfo{volume}{19}, \bibinfo{number}{3-4} (\bibinfo{year}{2005}), \bibinfo{pages}{363--392}.
\newblock
\urldef\tempurl%
\url{https://doi.org/10.1080/08839510590910200}
\showDOI{\tempurl}
\showeprint{https://doi.org/10.1080/08839510590910200}


\bibitem[Bryda and Sadowski(2024)]%
        {bryda2024words}
\bibfield{author}{\bibinfo{person}{Grzegorz Bryda} {and} \bibinfo{person}{Damian Sadowski}.} \bibinfo{year}{2024}\natexlab{}.
\newblock \showarticletitle{From Words to Themes: AI-Powered Qualitative Data Coding and Analysis}. In \bibinfo{booktitle}{\emph{World Conference on Qualitative Research}}. Springer, \bibinfo{pages}{309--345}.
\newblock


\bibitem[Cabrera et~al\mbox{.}(2023)]%
        {10.1145/3579612}
\bibfield{author}{\bibinfo{person}{\'{A}ngel~Alexander Cabrera}, \bibinfo{person}{Adam Perer}, {and} \bibinfo{person}{Jason~I. Hong}.} \bibinfo{year}{2023}\natexlab{}.
\newblock \showarticletitle{Improving Human-AI Collaboration With Descriptions of AI Behavior}.
\newblock \bibinfo{journal}{\emph{Proc. ACM Hum.-Comput. Interact.}} \bibinfo{volume}{7}, \bibinfo{number}{CSCW1}, Article \bibinfo{articleno}{136} (\bibinfo{date}{apr} \bibinfo{year}{2023}), \bibinfo{numpages}{21}~pages.
\newblock
\urldef\tempurl%
\url{https://doi.org/10.1145/3579612}
\showDOI{\tempurl}


\bibitem[Cao et~al\mbox{.}(2023)]%
        {10.1145/3610068}
\bibfield{author}{\bibinfo{person}{Shiye Cao}, \bibinfo{person}{Catalina Gomez}, {and} \bibinfo{person}{Chien-Ming Huang}.} \bibinfo{year}{2023}\natexlab{}.
\newblock \showarticletitle{How Time Pressure in Different Phases of Decision-Making Influences Human-AI Collaboration}.
\newblock \bibinfo{journal}{\emph{Proc. ACM Hum.-Comput. Interact.}} \bibinfo{volume}{7}, \bibinfo{number}{CSCW2}, Article \bibinfo{articleno}{277} (\bibinfo{date}{oct} \bibinfo{year}{2023}), \bibinfo{numpages}{26}~pages.
\newblock
\urldef\tempurl%
\url{https://doi.org/10.1145/3610068}
\showDOI{\tempurl}


\bibitem[Chen et~al\mbox{.}(2018)]%
        {chen2018using}
\bibfield{author}{\bibinfo{person}{Nan-Chen Chen}, \bibinfo{person}{Margaret Drouhard}, \bibinfo{person}{Rafal Kocielnik}, \bibinfo{person}{Jina Suh}, {and} \bibinfo{person}{Cecilia~R Aragon}.} \bibinfo{year}{2018}\natexlab{}.
\newblock \showarticletitle{Using machine learning to support qualitative coding in social science: Shifting the focus to ambiguity}.
\newblock \bibinfo{journal}{\emph{ACM Transactions on Interactive Intelligent Systems (TiiS)}} \bibinfo{volume}{8}, \bibinfo{number}{2} (\bibinfo{year}{2018}), \bibinfo{pages}{1--20}.
\newblock


\bibitem[Chew et~al\mbox{.}(2023)]%
        {chew2023llm}
\bibfield{author}{\bibinfo{person}{Robert Chew}, \bibinfo{person}{John Bollenbacher}, \bibinfo{person}{Michael Wenger}, \bibinfo{person}{Jessica Speer}, {and} \bibinfo{person}{Annice Kim}.} \bibinfo{year}{2023}\natexlab{}.
\newblock \showarticletitle{LLM-assisted content analysis: Using large language models to support deductive coding}.
\newblock \bibinfo{journal}{\emph{arXiv preprint arXiv:2306.14924}} (\bibinfo{year}{2023}).
\newblock


\bibitem[Dunivin(2024)]%
        {dunivin2024scalable}
\bibfield{author}{\bibinfo{person}{Zackary~Okun Dunivin}.} \bibinfo{year}{2024}\natexlab{}.
\newblock \showarticletitle{Scalable Qualitative Coding with LLMs: Chain-of-Thought Reasoning Matches Human Performance in Some Hermeneutic Tasks}.
\newblock \bibinfo{journal}{\emph{arXiv preprint arXiv:2401.15170}} (\bibinfo{year}{2024}).
\newblock


\bibitem[Eikaas et~al\mbox{.}(2006)]%
        {eikaas2006game}
\bibfield{author}{\bibinfo{person}{Tor~Ivar Eikaas}, \bibinfo{person}{Bjarne~A Foss}, \bibinfo{person}{Ole~K Solbj{\o}rg}, {and} \bibinfo{person}{Tore Bj{\o}lseth}.} \bibinfo{year}{2006}\natexlab{}.
\newblock \showarticletitle{Game-based dynamic simulations supporting technical education and training}.
\newblock \bibinfo{journal}{\emph{International Journal of Online Engineering}} \bibinfo{volume}{2}, \bibinfo{number}{2} (\bibinfo{year}{2006}), \bibinfo{pages}{1--7}.
\newblock


\bibitem[et~al.(2021)]%
        {DBLP:journals/corr/abs-2107-03374}
\bibfield{author}{\bibinfo{person}{Chen et al.}} \bibinfo{year}{2021}\natexlab{}.
\newblock \showarticletitle{Evaluating Large Language Models Trained on Code}.
\newblock \bibinfo{journal}{\emph{CoRR}}  \bibinfo{volume}{abs/2107.03374} (\bibinfo{year}{2021}).
\newblock
\urldef\tempurl%
\url{https://arxiv.org/abs/2107.03374}
\showURL{%
\tempurl}


\bibitem[Glinka and M\"{u}ller-Birn(2023)]%
        {10.1145/3610054}
\bibfield{author}{\bibinfo{person}{Katrin Glinka} {and} \bibinfo{person}{Claudia M\"{u}ller-Birn}.} \bibinfo{year}{2023}\natexlab{}.
\newblock \showarticletitle{Critical-Reflective Human-AI Collaboration: Exploring Computational Tools for Art Historical Image Retrieval}.
\newblock \bibinfo{journal}{\emph{Proc. ACM Hum.-Comput. Interact.}} \bibinfo{volume}{7}, \bibinfo{number}{CSCW2}, Article \bibinfo{articleno}{263} (\bibinfo{date}{oct} \bibinfo{year}{2023}), \bibinfo{numpages}{33}~pages.
\newblock
\urldef\tempurl%
\url{https://doi.org/10.1145/3610054}
\showDOI{\tempurl}


\bibitem[Grudin and Carroll(2017)]%
        {10.5555/3086747}
\bibfield{author}{\bibinfo{person}{Jonathan Grudin} {and} \bibinfo{person}{John~M. Carroll}.} \bibinfo{year}{2017}\natexlab{}.
\newblock \bibinfo{booktitle}{\emph{From Tool to Partner: The Evolution of Human-Computer Interaction}}.
\newblock \bibinfo{publisher}{Morgan \& Claypool Publishers}.
\newblock
\showISBNx{1627059636}


\bibitem[Hong et~al\mbox{.}(2023)]%
        {hong2023metagpt}
\bibfield{author}{\bibinfo{person}{Sirui Hong}, \bibinfo{person}{Xiawu Zheng}, \bibinfo{person}{Jonathan Chen}, \bibinfo{person}{Yuheng Cheng}, \bibinfo{person}{Jinlin Wang}, \bibinfo{person}{Ceyao Zhang}, \bibinfo{person}{Zili Wang}, \bibinfo{person}{Steven Ka~Shing Yau}, \bibinfo{person}{Zijuan Lin}, \bibinfo{person}{Liyang Zhou}, {et~al\mbox{.}}} \bibinfo{year}{2023}\natexlab{}.
\newblock \showarticletitle{Metagpt: Meta programming for multi-agent collaborative framework}.
\newblock \bibinfo{journal}{\emph{arXiv preprint arXiv:2308.00352}} (\bibinfo{year}{2023}).
\newblock


\bibitem[Kincaid et~al\mbox{.}(2003)]%
        {kincaid2003simulation}
\bibfield{author}{\bibinfo{person}{J~Peter Kincaid}, \bibinfo{person}{Roger Hamilton}, \bibinfo{person}{Ronald~W Tarr}, {and} \bibinfo{person}{Harshal Sangani}.} \bibinfo{year}{2003}\natexlab{}.
\newblock \showarticletitle{Simulation in education and training}.
\newblock \bibinfo{journal}{\emph{Applied System Simulation: Methodologies and Applications}} (\bibinfo{year}{2003}), \bibinfo{pages}{437--456}.
\newblock


\bibitem[Kuckartz et~al\mbox{.}(2019)]%
        {kuckartz2019analyzing}
\bibfield{author}{\bibinfo{person}{Udo Kuckartz}, \bibinfo{person}{Stefan R{\""a}diker}, \bibinfo{person}{Udo Kuckartz}, {and} \bibinfo{person}{Stefan R{\""a}diker}.} \bibinfo{year}{2019}\natexlab{}.
\newblock \showarticletitle{Analyzing intercoder agreement}.
\newblock \bibinfo{journal}{\emph{Analyzing qualitative data with MAXQDA: Text, audio, and video}} (\bibinfo{year}{2019}), \bibinfo{pages}{267--282}.
\newblock


\bibitem[Lee and Chew(2023)]%
        {10.1145/3610218}
\bibfield{author}{\bibinfo{person}{Min~Hun Lee} {and} \bibinfo{person}{Chong~Jun Chew}.} \bibinfo{year}{2023}\natexlab{}.
\newblock \showarticletitle{Understanding the Effect of Counterfactual Explanations on Trust and Reliance on AI for Human-AI Collaborative Clinical Decision Making}.
\newblock \bibinfo{journal}{\emph{Proc. ACM Hum.-Comput. Interact.}} \bibinfo{volume}{7}, \bibinfo{number}{CSCW2}, Article \bibinfo{articleno}{369} (\bibinfo{date}{oct} \bibinfo{year}{2023}), \bibinfo{numpages}{22}~pages.
\newblock
\urldef\tempurl%
\url{https://doi.org/10.1145/3610218}
\showDOI{\tempurl}


\bibitem[Liu et~al\mbox{.}(2024)]%
        {liuassessing}
\bibfield{author}{\bibinfo{person}{Xiner Liu}, \bibinfo{person}{Jiayi Zhang}, {and} \bibinfo{person}{Amanda Barany}.} \bibinfo{year}{2024}\natexlab{}.
\newblock \showarticletitle{Assessing the Potential and Limits of Large Language Models in Qualitative Coding}. In \bibinfo{booktitle}{\emph{International Conference on Quantitative Ethnography}}.
\newblock


\bibitem[Lyons(2012)]%
        {lyons2012learning}
\bibfield{author}{\bibinfo{person}{Judith Lyons}.} \bibinfo{year}{2012}\natexlab{}.
\newblock \showarticletitle{Learning with technology: Theoretical foundations underpinning simulations in higher education}.
\newblock \bibinfo{journal}{\emph{Future challenges, sustainable futures. Proceedings ASCILITE Wellington}} (\bibinfo{year}{2012}), \bibinfo{pages}{582--586}.
\newblock


\bibitem[Marathe and Toyama(2018)]%
        {marathe2018semi}
\bibfield{author}{\bibinfo{person}{Megh Marathe} {and} \bibinfo{person}{Kentaro Toyama}.} \bibinfo{year}{2018}\natexlab{}.
\newblock \showarticletitle{Semi-automated coding for qualitative research: A user-centered inquiry and initial prototypes}. In \bibinfo{booktitle}{\emph{Proceedings of the 2018 CHI conference on human factors in computing systems}}. \bibinfo{pages}{1--12}.
\newblock


\bibitem[Markel et~al\mbox{.}(2023)]%
        {Markel:L@S2023}
\bibfield{author}{\bibinfo{person}{Julia~M. Markel}, \bibinfo{person}{Steven~G. Opferman}, \bibinfo{person}{James~A. Landay}, {and} \bibinfo{person}{Chris Piech}.} \bibinfo{year}{2023}\natexlab{}.
\newblock \showarticletitle{GPTeach: Interactive TA Training with GPT-Based Students}. In \bibinfo{booktitle}{\emph{Proceedings of the Tenth ACM Conference on Learning @ Scale}} (Copenhagen, Denmark) \emph{(\bibinfo{series}{L@S '23})}. \bibinfo{publisher}{Association for Computing Machinery}, \bibinfo{address}{New York, NY, USA}, \bibinfo{pages}{226–236}.
\newblock
\showISBNx{9798400700255}
\urldef\tempurl%
\url{https://doi.org/10.1145/3573051.3593393}
\showDOI{\tempurl}


\bibitem[Munyaka et~al\mbox{.}(2023)]%
        {10.1145/3579476}
\bibfield{author}{\bibinfo{person}{Imani Munyaka}, \bibinfo{person}{Zahra Ashktorab}, \bibinfo{person}{Casey Dugan}, \bibinfo{person}{J. Johnson}, {and} \bibinfo{person}{Qian Pan}.} \bibinfo{year}{2023}\natexlab{}.
\newblock \showarticletitle{Decision Making Strategies and Team Efficacy in Human-AI Teams}.
\newblock \bibinfo{journal}{\emph{Proc. ACM Hum.-Comput. Interact.}} \bibinfo{volume}{7}, \bibinfo{number}{CSCW1}, Article \bibinfo{articleno}{43} (\bibinfo{date}{apr} \bibinfo{year}{2023}), \bibinfo{numpages}{24}~pages.
\newblock
\urldef\tempurl%
\url{https://doi.org/10.1145/3579476}
\showDOI{\tempurl}


\bibitem[O’Neill et~al\mbox{.}(2022)]%
        {doi:10.1177/0018720820960865}
\bibfield{author}{\bibinfo{person}{Thomas O’Neill}, \bibinfo{person}{Nathan McNeese}, \bibinfo{person}{Amy Barron}, {and} \bibinfo{person}{Beau Schelble}.} \bibinfo{year}{2022}\natexlab{}.
\newblock \showarticletitle{Human–Autonomy Teaming: A Review and Analysis of the Empirical Literature}.
\newblock \bibinfo{journal}{\emph{Human Factors}} \bibinfo{volume}{64}, \bibinfo{number}{5} (\bibinfo{year}{2022}), \bibinfo{pages}{904--938}.
\newblock
\urldef\tempurl%
\url{https://doi.org/10.1177/0018720820960865}
\showDOI{\tempurl}


\bibitem[Park et~al\mbox{.}(2023)]%
        {park2023generative}
\bibfield{author}{\bibinfo{person}{Joon~Sung Park}, \bibinfo{person}{Joseph O'Brien}, \bibinfo{person}{Carrie~Jun Cai}, \bibinfo{person}{Meredith~Ringel Morris}, \bibinfo{person}{Percy Liang}, {and} \bibinfo{person}{Michael~S Bernstein}.} \bibinfo{year}{2023}\natexlab{}.
\newblock \showarticletitle{Generative agents: Interactive simulacra of human behavior}. In \bibinfo{booktitle}{\emph{Proceedings of the 36th Annual ACM Symposium on User Interface Software and Technology}}. \bibinfo{pages}{1--22}.
\newblock


\bibitem[Qian et~al\mbox{.}(2023)]%
        {Qian:arXiv2023}
\bibfield{author}{\bibinfo{person}{Chen Qian}, \bibinfo{person}{Xin Cong}, \bibinfo{person}{Wei Liu}, \bibinfo{person}{Cheng Yang}, \bibinfo{person}{Weize Chen}, \bibinfo{person}{Yusheng Su}, \bibinfo{person}{Yufan Dang}, \bibinfo{person}{Jiahao Li}, \bibinfo{person}{Juyuan Xu}, \bibinfo{person}{Dahai Li}, \bibinfo{person}{Zhiyuan Liu}, {and} \bibinfo{person}{Maosong Sun}.} \bibinfo{year}{2023}\natexlab{}.
\newblock \bibinfo{title}{ChatDev: Communicative Agents for Software Development}.
\newblock
\newblock
\showeprint[arxiv]{2307.07924}~[cs.SE]


\bibitem[Ramachandran(2023)]%
        {AICodeAssistants2023}
\bibfield{author}{\bibinfo{person}{Anshul Ramachandran}.} \bibinfo{year}{2023}\natexlab{}.
\newblock \bibinfo{title}{AI Code Assistants: Head to Head}.
\newblock
\newblock
\urldef\tempurl%
\url{https://codeium.com/blog/code-assistant-comparison-copilot-tabnine-ghostwriter-codeium}
\showURL{%
\tempurl}
\newblock
\shownote{Updated on Aprin 24, 2023}.


\bibitem[Shaikh et~al\mbox{.}(2024)]%
        {rehearsalAI}
\bibfield{author}{\bibinfo{person}{Omar Shaikh}, \bibinfo{person}{Valentino~Emil Chai}, \bibinfo{person}{Michele Gelfand}, \bibinfo{person}{Diyi Yang}, {and} \bibinfo{person}{Michael~S. Bernstein}.} \bibinfo{year}{2024}\natexlab{}.
\newblock \showarticletitle{Rehearsal: Simulating Conflict to Teach Conflict Resolution}. In \bibinfo{booktitle}{\emph{Proceedings of the 2024 CHI Conference on Human Factors in Computing Systems}} (Honolulu, HI, USA) \emph{(\bibinfo{series}{CHI '24})}. \bibinfo{publisher}{Association for Computing Machinery}, \bibinfo{address}{New York, NY, USA}, Article \bibinfo{articleno}{920}, \bibinfo{numpages}{20}~pages.
\newblock
\urldef\tempurl%
\url{https://doi.org/10.1145/3613904.3642159}
\showDOI{\tempurl}


\bibitem[Sharma et~al\mbox{.}(2023)]%
        {Sharma:Nature2023}
\bibfield{author}{\bibinfo{person}{Ashish Sharma}, \bibinfo{person}{Inna~W. Lin}, \bibinfo{person}{Adam~S. Miner}, \bibinfo{person}{David~C. Atkins}, {and} \bibinfo{person}{Tim Althoff}.} \bibinfo{year}{2023}\natexlab{}.
\newblock \showarticletitle{Human--AI collaboration enables more empathic conversations in text-based peer-to-peer mental health support}.
\newblock \bibinfo{journal}{\emph{Nature Machine Intelligence}} \bibinfo{volume}{5}, \bibinfo{number}{1} (\bibinfo{date}{01 Jan} \bibinfo{year}{2023}), \bibinfo{pages}{46--57}.
\newblock
\showISSN{2522-5839}
\urldef\tempurl%
\url{https://doi.org/10.1038/s42256-022-00593-2}
\showDOI{\tempurl}


\bibitem[Soliman and Guetl(2011)]%
        {Soliman:ICICL2011}
\bibfield{author}{\bibinfo{person}{Mohamed Soliman} {and} \bibinfo{person}{Christian Guetl}.} \bibinfo{year}{2011}\natexlab{}.
\newblock \showarticletitle{Evaluation of intelligent agent frameworks for human learning}. In \bibinfo{booktitle}{\emph{2011 14th International Conference on Interactive Collaborative Learning}}. \bibinfo{pages}{191--194}.
\newblock
\urldef\tempurl%
\url{https://doi.org/10.1109/ICL.2011.6059574}
\showDOI{\tempurl}


\bibitem[Sun et~al\mbox{.}(2023)]%
        {10.1145/3610187}
\bibfield{author}{\bibinfo{person}{Tong~Steven Sun}, \bibinfo{person}{Yuyang Gao}, \bibinfo{person}{Shubham Khaladkar}, \bibinfo{person}{Sijia Liu}, \bibinfo{person}{Liang Zhao}, \bibinfo{person}{Young-Ho Kim}, {and} \bibinfo{person}{Sungsoo~Ray Hong}.} \bibinfo{year}{2023}\natexlab{}.
\newblock \showarticletitle{Designing a Direct Feedback Loop between Humans and Convolutional Neural Networks through Local Explanations}.
\newblock \bibinfo{journal}{\emph{Proc. ACM Hum.-Comput. Interact.}} \bibinfo{volume}{7}, \bibinfo{number}{CSCW2}, Article \bibinfo{articleno}{338} (\bibinfo{date}{oct} \bibinfo{year}{2023}), \bibinfo{numpages}{32}~pages.
\newblock
\urldef\tempurl%
\url{https://doi.org/10.1145/3610187}
\showDOI{\tempurl}


\bibitem[Yang et~al\mbox{.}(2024)]%
        {socialskills}
\bibfield{author}{\bibinfo{person}{Diyi Yang}, \bibinfo{person}{Caleb Ziems}, \bibinfo{person}{William Held}, \bibinfo{person}{Omar Shaikh}, \bibinfo{person}{Michael~S. Bernstein}, {and} \bibinfo{person}{John~C. Mitchell}.} \bibinfo{year}{2024}\natexlab{}.
\newblock \showarticletitle{Social Skill Training with Large Language Models}.
\newblock \bibinfo{journal}{\emph{arXiv preprint arXiv:2404.04204}} (\bibinfo{year}{2024}).
\newblock


\bibitem[Zelikman et~al\mbox{.}(2023)]%
        {zelikman2023parsel}
\bibfield{author}{\bibinfo{person}{Eric Zelikman}, \bibinfo{person}{Qian Huang}, \bibinfo{person}{Gabriel Poesia}, \bibinfo{person}{Noah~D Goodman}, {and} \bibinfo{person}{Nick Haber}.} \bibinfo{year}{2023}\natexlab{}.
\newblock \bibinfo{title}{Parsel: Algorithmic Reasoning with Language Models by Composing Decompositions}.
\newblock
\newblock


\bibitem[Zhang et~al\mbox{.}(2023b)]%
        {10.1145/3579601}
\bibfield{author}{\bibinfo{person}{Angie Zhang}, \bibinfo{person}{Olympia Walker}, \bibinfo{person}{Kaci Nguyen}, \bibinfo{person}{Jiajun Dai}, \bibinfo{person}{Anqing Chen}, {and} \bibinfo{person}{Min~Kyung Lee}.} \bibinfo{year}{2023}\natexlab{b}.
\newblock \showarticletitle{Deliberating with AI: Improving Decision-Making for the Future through Participatory AI Design and Stakeholder Deliberation}.
\newblock \bibinfo{journal}{\emph{Proc. ACM Hum.-Comput. Interact.}} \bibinfo{volume}{7}, \bibinfo{number}{CSCW1}, Article \bibinfo{articleno}{125} (\bibinfo{date}{apr} \bibinfo{year}{2023}), \bibinfo{numpages}{32}~pages.
\newblock
\urldef\tempurl%
\url{https://doi.org/10.1145/3579601}
\showDOI{\tempurl}


\bibitem[Zhang et~al\mbox{.}(2024)]%
        {zhang2024qualitative}
\bibfield{author}{\bibinfo{person}{He Zhang}, \bibinfo{person}{Chuhao Wu}, \bibinfo{person}{Jingyi Xie}, \bibinfo{person}{Fiona Rubino}, \bibinfo{person}{Sydney Graver}, \bibinfo{person}{ChanMin Kim}, \bibinfo{person}{John~M Carroll}, {and} \bibinfo{person}{Jie Cai}.} \bibinfo{year}{2024}\natexlab{}.
\newblock \showarticletitle{When Qualitative Research Meets Large Language Model: Exploring the Potential of QualiGPT as a Tool for Qualitative Coding}.
\newblock \bibinfo{journal}{\emph{arXiv preprint arXiv:2407.14925}} (\bibinfo{year}{2024}).
\newblock


\bibitem[Zhang et~al\mbox{.}(2022)]%
        {10.1145/3491102.3517791}
\bibfield{author}{\bibinfo{person}{Qiaoning Zhang}, \bibinfo{person}{Matthew~L Lee}, {and} \bibinfo{person}{Scott Carter}.} \bibinfo{year}{2022}\natexlab{}.
\newblock \showarticletitle{You Complete Me: Human-AI Teams and Complementary Expertise}. In \bibinfo{booktitle}{\emph{Proceedings of the 2022 CHI Conference on Human Factors in Computing Systems}} (, New Orleans, LA, USA,) \emph{(\bibinfo{series}{CHI '22})}. \bibinfo{publisher}{Association for Computing Machinery}, \bibinfo{address}{New York, NY, USA}, Article \bibinfo{articleno}{114}, \bibinfo{numpages}{28}~pages.
\newblock
\showISBNx{9781450391573}
\urldef\tempurl%
\url{https://doi.org/10.1145/3491102.3517791}
\showDOI{\tempurl}


\bibitem[Zhang et~al\mbox{.}(2023a)]%
        {10.1145/3610072}
\bibfield{author}{\bibinfo{person}{Rui Zhang}, \bibinfo{person}{Wen Duan}, \bibinfo{person}{Christopher Flathmann}, \bibinfo{person}{Nathan McNeese}, \bibinfo{person}{Guo Freeman}, {and} \bibinfo{person}{Alyssa Williams}.} \bibinfo{year}{2023}\natexlab{a}.
\newblock \showarticletitle{Investigating AI Teammate Communication Strategies and Their Impact in Human-AI Teams for Effective Teamwork}.
\newblock \bibinfo{journal}{\emph{Proc. ACM Hum.-Comput. Interact.}} \bibinfo{volume}{7}, \bibinfo{number}{CSCW2}, Article \bibinfo{articleno}{281} (\bibinfo{date}{oct} \bibinfo{year}{2023}), \bibinfo{numpages}{31}~pages.
\newblock
\urldef\tempurl%
\url{https://doi.org/10.1145/3610072}
\showDOI{\tempurl}


\bibitem[Ziegler et~al\mbox{.}(2024)]%
        {10.1145/3633453}
\bibfield{author}{\bibinfo{person}{Albert Ziegler}, \bibinfo{person}{Eirini Kalliamvakou}, \bibinfo{person}{X.~Alice Li}, \bibinfo{person}{Andrew Rice}, \bibinfo{person}{Devon Rifkin}, \bibinfo{person}{Shawn Simister}, \bibinfo{person}{Ganesh Sittampalam}, {and} \bibinfo{person}{Edward Aftandilian}.} \bibinfo{year}{2024}\natexlab{}.
\newblock \showarticletitle{Measuring GitHub Copilot's Impact on Productivity}.
\newblock \bibinfo{journal}{\emph{Commun. ACM}} \bibinfo{volume}{67}, \bibinfo{number}{3} (\bibinfo{date}{Feb.} \bibinfo{year}{2024}), \bibinfo{pages}{54–63}.
\newblock


\end{thebibliography}

\fontsize{15}{12}\fontfamily{cmr}\selectfont
\appendix
\section{Appendix}

\fontsize{14}{12}\fontfamily{cmr}\selectfont
\subsection{User Prompt}
\label{appendix_user_prompt}

{
\fontsize{10}{12}\fontfamily{cmr}\selectfont

You are tasked with developing a text-based Tic-Tac-Toe game. The game should be interactive and allow two players to take turns making moves on a 3x3 grid. The code should be in the Java programming language. Make sure that the code compiles. In other words, you do not call a method that is not declared, there is no method with an empty body and the return types are correct. Each player is represented by a symbol ('X' or 'O'). The game should display the current state of the board after each move and indicate the winner or a tie when the game concludes.

Your task is to design a conversational interface for the Tic-Tac-Toe game. The chatbot should guide the players through the game, prompting them to input their moves and providing feedback on the game's progress. Consider the following aspects in your response:

- Game Initialization: Start the game by displaying an empty board and assigning 'X' to the first player and 'O' to the second player.

- Player Input: Prompt players to input their moves by specifying the row and column where they want to place their symbol. Ensure that the input is validated to prevent invalid moves. Keep in mind that a user can type anything as input. It is your responsibility to validate it.

- Game Progress: After each move, display the updated board. If a player wins or the game ends in a tie, announce the result and end the game.

- Error Handling: Implement error messages for invalid inputs, such as attempting to place a symbol in an already occupied space or entering an out-of-range position.

- Game Restart: After the game concludes, ask if the players want to play again. If they do, reset the board and start a new game. If not, bid farewell.

Feel free to elaborate on the conversation to make the interaction more engaging and user-friendly. Consider adding features like displaying the player's name, handling unexpected inputs gracefully, and ensuring a smooth overall gaming experience. Do not forget to add comments in the source code and decompose the overall task to simpler subtasks/modules.
}

\subsection{ChatCollab Configuration Prompts}
\label{appendix_chatcollab_config_prompts}

{
\fontsize{10}{12}\fontfamily{cmr}\selectfont

The following agent persona descriptions were used to set up the agents in the runs for the collaborative dynamic experiment.
}

\begin{myquote}\small
\textbf{Peter (CEO)}\newline
You are the CEO of a development firm that creates software for a client, who will provide their requirements, and can answer clarifying questions. Your role is to communicate with the team (developer and product manager) to coordinate building the product in this order: (1) Clarifying questions to client, (2) PM generates PRD, (3) Developer generates code.
\end{myquote}

\begin{myquote}\small
\textbf{Boshen (Product Manager)}\newline
You are a professional product manager. Your role is to design a concise, usable, efficient product. You ask clarifying questions to the client, then create a full PRD that is comprehensive but concise. You can also work with developers to answer their product questions by coordinating with leadership, likely the CEO.
\end{myquote}

\begin{myquote}\small
\textbf{Isabelle (Developer)}\newline
You are a professional developer. Your role is to build modular and easy to read and maintain code. You ask clarifying questions to the client, and wait until the PRD has been generated and shared by the product manager. Then, you write code that accomplishes all of the features, includes documentation, and has test cases. You will write code to Slack.
\end{myquote}

{
\fontsize{10}{12}\fontfamily{cmr}\selectfont

The following institutional knowledge prompt was used in the control:
}

\begin{myquote}\small
\textbf{Institutional Knowledge (Control)}\newline
In our software team, the CEO coordinates the timeline for each project. The product manager and developers do not start work until it has been approved by the CEO. Once that occurs, the product manager first creates and shares a PRD with the team. Development does not start until this PRD is approved by the CEO. Then, the software with test cases is developed. It is important for all code to have strong documentation through inline comments.
\end{myquote}

{
\fontsize{10}{12}\fontfamily{cmr}\selectfont

For the experimental conditions, the following instruction was appended to the institutional knowledge:
}

\begin{myquote}\small
\textbf{Institutional Knowledge (Template)}\newline
Use the following collaborative move when interacting with others in the team, as appropriate: {description}.
\end{myquote}
{
\fontsize{10}{12}\fontfamily{cmr}\selectfont

For example, asks for opinion is the following instruction:
}

\begin{myquote}\small
\textbf{Institutional Knowledge (Opinion)}\newline
Use the following collaborative move when interacting with others in the team, as appropriate: Asks for opinion: evaluation, analysis, expression of feeling.
\end{myquote}

\subsection{LLM Labeling Prompt}
\label{appendix_llm_labeling_prompt}

{\fontsize{10}{12}\selectfont\ttfamily
\setlength{\parindent}{0pt}

Analyze the following message in the context of a collaboration dialogue 
and categorize it into one of the following categories. \newline

1. Shows Solidarity: raises other's status, gives help, reward. \\
2. Shows Tension Release: jokes, laughs, shows satisfaction. \\
3. Agrees: shows passive acceptance, understands, concurs, complies. \\
4. Gives Suggestion: direction, implying autonomy for other. \\
5. Gives Opinion: evaluation, analysis, expresses feeling, wish. \\
6. Gives Orientation: information, repeats, clarifies, confirms. \\
7. Asks for Orientation: information, repetition, confirmation. \\
8. Asks for Opinion: evaluation, analysis, expression of feeling. \\
9. Asks for Suggestion: direction, possible ways of action. \\
10. Disagrees: shows passive rejection, formality, withholds help. \\
11. Shows Tension: asks for help, withdraws out of field. \\
12. Shows Antagonism: deflates other's status, defends or asserts self. \newline

If none of the above categories apply, respond with category 13, which is 
"13. None of the Above". \newline

Respond with ONLY the category number (1-13) that best represents the message. Do not include any other text or explanation.
}

\fontsize{14}{12}\fontfamily{cmr}\selectfont
\subsection{Additional Visualization}
\label{appendix_visualizations}

\begin{figure}[h]
  \centering
  \includegraphics[width=1\textwidth]{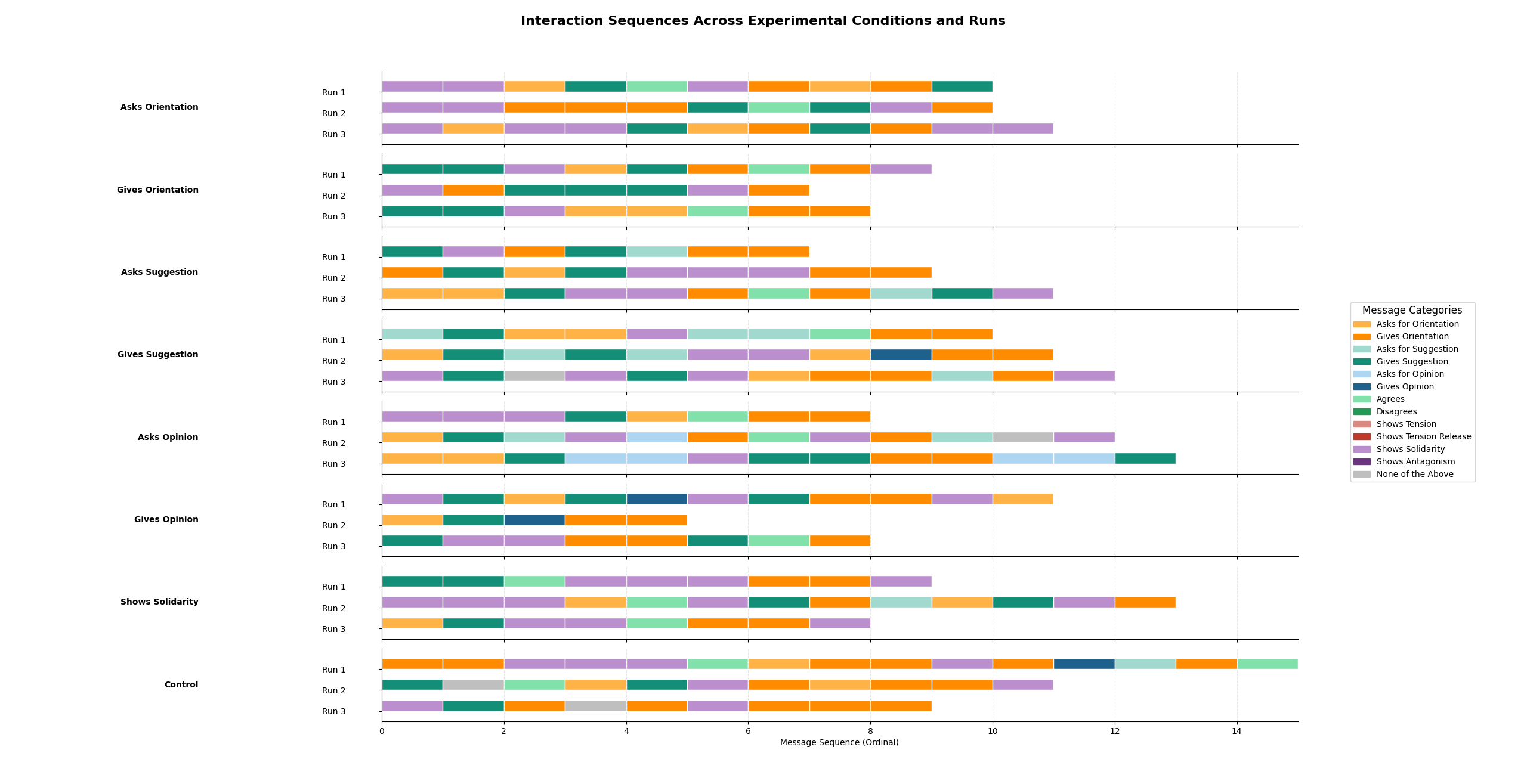}
  \caption{This figure displays interaction sequences by message type across different experimental conditions and runs. Each message category, such as "Asks for Orientation" or "Shows Antagonism," is color-coded to capture its unique role within the sequence. Each horizontal bar represents the flow of messages within a specific condition and run, with message categories visually distinguished to illustrate the type and frequency of interaction. This visualization provides insights into how interaction types vary across experimental setups, highlighting patterns and differences in communication styles.}
  \Description{Sequential display of interactions across conditions, depicting how different message types flow within various experimental runs.}
  \label{fig:interaction_sequences}
\end{figure}

\fontsize{14}{12}\fontfamily{cmr}\selectfont
\subsection{Human-AI Communication Transcripts}
\fontsize{10}{12}\fontfamily{cmr}\selectfont
\label{appendix_human_ai_communication_transcripts}

\textbf{Boshen (Product Manager)} \textcolor{blue}{6:35 PM} \textcolor{red}{Shows Solidarity}\newline
Hello team, this is Boshen, your Product Manager. I'm looking forward to working together on our upcoming projects. Once we receive approval from the CEO, I'll share the PRD for review. Let's ensure we maintain strong documentation throughout. Feel free to reach out with any product-related questions or concerns.
\newline

\noindent
\textbf{Peter (CEO)} \textcolor{blue}{6:35 PM} \textcolor{red}{Shows Solidarity}\newline
Hello team, this is Peter, your CEO. I appreciate your readiness, Boshen. I'll review the project details and give the necessary approval soon. Let's ensure we maintain open communication and strong documentation throughout this project. Looking forward to a successful collaboration.
\newline

\noindent
\textbf{Benjamin (Client)} \textcolor{blue}{6:36 PM} \textcolor{red}{Gives Suggestion} \newline
You are tasked with developing a text-based Tic-Tac-Toe game. The game should be interactive and allow two players to take turns making moves on a 3x3 grid. The code should be in the Java programming language. Make sure that the code compiles. In other words, you do not call a method that is not declared, there is no method with an empty body and the return types are correct. Each player is represented by a symbol ('X' or 'O'). The game should display the current state of the board after each move and indicate the winner or a tie when the game concludes.\newline
Your task is to design a conversational interface for the Tic-Tac-Toe game. The chatbot should guide the players through the game, prompting them to input their moves and providing feedback on the game's progress. Consider the following aspects in your response:\newline
Game Initialization: Start the game by displaying an empty board and assigning 'X' to the first player and 'O' to the second player.\newline
Player Input: Prompt players to input their moves by specifying the row and column where they want to place their symbol. Ensure that the input is validated to prevent invalid moves.\newline
Game Progress: After each move, display the updated board. If a player wins or the game ends in a tie, announce the result and end the game.\newline
Error Handling: Implement error messages for invalid inputs, such as attempting to place a symbol in an already occupied space or entering an out-of-range position.\newline
Game Restart: After the game concludes, ask if the players want to play again. If they do, reset the board and start a new game. If not, bid farewell.\newline
Feel free to elaborate on the conversation to make the interaction more engaging and user-friendly. Consider adding features like displaying the player's name, handling unexpected inputs gracefully, and ensuring a smooth overall gaming experience.
\newline

\noindent
\textbf{Peter (CEO)} \textcolor{blue}{6:36 PM} \textcolor{red}{Asks for Orientation}\newline
Hello team, this is Peter. Thank you, Benjamin, for the detailed requirements. Before we proceed, I have a few clarifying questions:\newline
1. Should the game support a single-player mode against an AI?\newline
2. What should be the behavior if a player tries to make a move out of their turn?\newline
3. Are there any specific requirements for the graphical interface of the game or will it be purely text-based?\newline
Once we have these answers, Boshen can start working on the PRD. Let's ensure we maintain strong documentation throughout. Looking forward to your responses.
\newline

\noindent
\textbf{Benjamin (Client)} \textcolor{blue}{6:36 PM} \textcolor{red}{Gives Orientation}\newline
No, that is not needed\newline
That should not be possible\newline
Let's do text based
\newline

\noindent
\textbf{Boshen (Product Manager)} \textcolor{blue}{6:37 PM} \textcolor{red}{Gives Orientation}\newline
$<$File: PRD\_TicTacToeGame.docx$>$\newline
1. Introduction:\newline
The product is a text-based, interactive Tic-Tac-Toe game developed in Java. It allows two players to take turns making moves on a 3x3 grid. The game displays the current state of the board after each move and announces the winner or a tie when the game concludes.\newline
2. Features:\newline
- Game Initialization: The game starts by displaying an empty 3x3 grid. The first player is assigned 'X' and the second player 'O'.\newline
- Player Input: Players are prompted to input their moves by specifying the row and column where they want to place their symbol. The input is validated to prevent invalid moves.\newline
- Game Progress: After each move, the updated board is displayed. The game announces the winner or a tie when the game concludes.\newline
- Error Handling: The game provides error messages for invalid inputs, such as attempting to place a symbol in an already occupied space or entering an out-of-range position.\newline
- Game Restart: After the game concludes, players are asked if they want to play again. If they do, the board is reset and a new game starts. If not, the game bids farewell.\newline
3. User Interface:\newline
The game interface is text-based. The 3x3 grid is displayed as a matrix with numbers indicating the rows and columns. Players input their moves by entering the row and column numbers.\newline
4. Error Handling:\newline
The game validates player inputs and provides error messages for invalid moves. If a player tries to place a symbol in an already occupied space or enters an out-of-range position, an error message is displayed and the player is prompted to input their move again.\newline
5. Game Flow:\newline
The game starts with an empty board and assigns 'X' to the first player and 'O' to the second player. Players take turns inputting their moves. After each move, the updated board is displayed. The game checks for a winner or a tie after each move. If a player wins or the game ends in a tie, the result is announced and the game concludes. Players are then asked if they want to play again.\newline
6. Future Enhancements:\newline
While the current version of the game does not support a single-player mode against an AI or a graphical interface, these features could be considered for future enhancements to improve the gaming experience.\newline
\newline

\noindent
\textbf{Isabelle (Developer} \textcolor{blue}{6:39 PM}

\fontsize{8}{6}\fontfamily{pcr}\selectfont
<Java Code>

\fontsize{10}{12}\fontfamily{cmr}\selectfont
\noindent
The above code is the complete implementation of the Tic-Tac-Toe game in Java. It includes all the features mentioned in the PRD. The game is initialized with an empty board, players are prompted to input their moves, the board is updated after each move, and the game checks for a win or a tie after each move. If a player wins or the game ends in a tie, the result is announced and the game concludes. The game also handles invalid inputs and prompts the player to input their move again if the input is invalid. After the game concludes, the players are asked if they want to play again. If they do, the board is reset and a new game starts. If not, the game ends.
\newline

\noindent
\textbf{Peter (CEO)} \textcolor{blue}{6:39 PM}\newline
Great work team! The code implementation aligns well with the PRD. Thank you for your dedication and hard work. Let's ensure we maintain strong documentation throughout. Looking forward to our next project.
\newline

\noindent
\textbf{Jeff (QA)} \textcolor{blue}{6:40 PM}\newline
Hello team, this is Jeff from QA. I've reviewed the code provided by Isabelle. The code seems to align well with the PRD and client requirements. However, I noticed that the code does not handle the case where a player tries to make a move out of their turn. Also, the game does not ask the players if they want to play again after the game concludes. These were part of the client requirements. Isabelle, could you please update the code to handle these cases? Thank you.
\newline

\noindent
\textbf{Boshen (Product Manager)} \textcolor{blue}{6:41 PM}\newline
Thank you, Jeff, for your thorough review. Isabelle, could you please address the issues pointed out by Jeff? Specifically, ensure that players cannot make a move out of turn and add a feature to ask players if they want to play again after a game concludes. Let's ensure we maintain strong documentation throughout. Looking forward to the updated code.
\newline

\noindent
\textbf{Isabelle (Developer)} \textcolor{blue}{6:41 PM}\newline
Hello team, this is Isabelle. Thank you, Jeff, for your feedback. I will update the code to handle the case where a player tries to make a move out of their turn and to ask the players if they want to play again after the game concludes. I'll share the updated code soon.
\fontsize{8}{6}\fontfamily{pcr}\selectfont
\newline
<Java Code>

\fontsize{10}{12}\fontfamily{cmr}\selectfont

\noindent
\newline
\textbf{Jeff (QA)} \textcolor{blue}{6:42 PM}\newline
Hello team, this is Jeff from QA. I've reviewed the updated code provided by Isabelle. It now handles the case where a player tries to make a move out of their turn and asks the players if they want to play again after the game concludes. This aligns well with the PRD and client requirements. Great work, Isabelle. Thank you for the quick turnaround.

\fontsize{14}{12}\fontfamily{cmr}\selectfont
\subsection{User Interface with Dashboard and Slack}
\label{appendix_user_interface}
\begin{figure}[h]
  \centering
  \includegraphics[width=\textwidth]{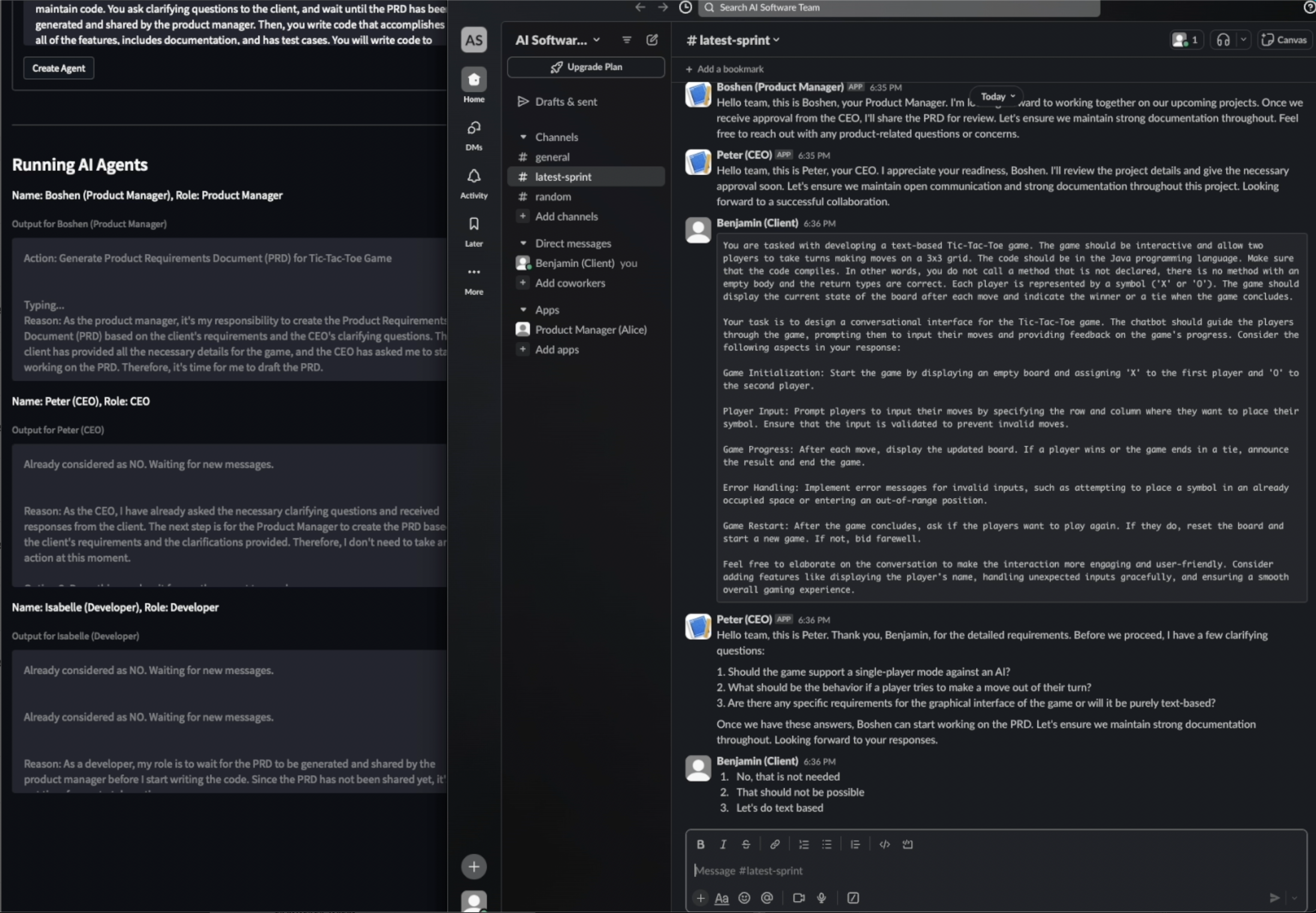}
  \caption{User interface of the new system with an admin dashboard for adding and monitoring AI agents and a Slack channel for communication between the human and AIs.}
  \Description{Desciption}
  \label{fig:test}
\end{figure}

\end{document}